\documentclass[usenatbib]{mn2e}
\usepackage{graphicx}
\usepackage{subfigure}
\usepackage{amssymb}
\usepackage{cancel}
\usepackage{bm}
\usepackage[draft]{hyperref}

\newcommand{\angstrom}{\mbox{\normalfont\AA}}

\title[Modelling the star-disc fly-by in RW Aurigae]{A tidal encounter caught in the act: modelling a star-disc fly-by in the young RW Aurigae system}
\author[F. Dai et al.]{Fei Dai$^{1,2}$\thanks{E-mail:
fd284@mit.edu}, Stefano Facchini$^{2}$\thanks{E-mail: facchini@ast.cam.ac.uk}, Cathie J. Clarke$^{2}$ and Thomas J. Haworth$^{2}$\\
$^{1}$ Kavli Institute for Astrophysics and Space Research, Massachusetts Institute of Technology, Cambridge, MA 02139, USA\\
$^{2}$ Institute of Astronomy, Madingley Rd, Cambridge, CB3 0HA, UK}
\begin{document}

\date{Accepted ???. Received ???; in original form ???}

\pagerange{\pageref{firstpage}--\pageref{lastpage}} \pubyear{2014}

\maketitle
\label{firstpage}

\begin{abstract}
RW Aurigae (RW Aur) is a binary star system with a long molecular arm trailing the primary star. Cabrit et al. (2006) noted the resemblance between this extended structure and the
tidal arm stripped from the primary star
in the simulations of star-disc encounters by Clarke \& Pringle (1993). In this paper we use new hydrodynamical models and synthetic observations
to fit many of the parameters of RW Aur. Using
hydrodynamic models we find that the morphological appearance of 
RW Aur can be indeed explained by a tidal encounter with the secondary star. We reproduce all the major morphological and kinematic features of the system. Using radiative transfer calculations, we find that synthetic CO and dust continuum observations of our hydrodynamic models agree well with observations. We reproduce all the main features
of the line profiles, from emission fluxes to the optical depth of the different components of the system. The agreement between observations and simulations thus lends strong support to the hypothesis of a tidal encounter scenario. Finally, we propose a possible solution for the origin of the dimming of the primary star observed in 2010/2011 by Rodriguez et al. (2013).

\end{abstract}


\section{Introduction}
\label{introduction}
 RW Aurigae is a  young stellar system comprising two roughly
solar mass stars at a projected separation of $1.5\arcsec$, i.e. $200$\,AU at a distance $d\simeq140$\,pc \citep[cfr. Hipparcos catalogue,][]{2007A&A...474..653V}. The
primary is a Classical T Tauri Star which is characterized by
an unusually high ratio of accretion rate \citep[$2-10 \times 10^{-7}
M_\odot$ yr$^{-1}$,][]{1989ApJ...341..340B, 1995ApJ...452..736H} to
disc mass \citep[$\sim10^{-4} M_\odot$ as derived from mm continuum dust emission,][]{
1995ApJ...439..288O}.

  The most notable feature of the RW Aurigae system is the $600$ AU
long arm structure trailing from the disc of RW Aur A which was imaged
by \citet{2006A&A...452..897C} in $^{12}$CO$(J=2-1)$ and $^{12}$CO$(J=1-0)$ with the IRAM(Institut de Radioastronomie Millimétrique) Plateau de Bure Interferometer.
\citet{2006A&A...452..897C} noted the resemblance between this structure and the
tidal arm stripped from the primary star
in the simulations of star-disc encounters by \citet{1993MNRAS.261..190C}. 
Although these
simulations were not designed to fit a particular observation,
the superficial resemblance between the morphology of RW Aur and the
prograde encounter of \citet{1993MNRAS.261..190C} was striking.
It is however unclear without detailed modelling whether simulations
can fit all the parameters of the system, particularly the gas kinematics
and the available radial velocity and proper motion data for the two stars.
Moreover \citet{2012ARep...56..686B} presented evidence from proper motion data
that the encounter might in fact be retrograde in which case the
\citet{1993MNRAS.261..190C} models would predict a morphology unlike that observed
in RW Aur.
\citet{2012ARep...56..686B} have on
this basis proposed an alternative model involving circumbinary disc
accretion. In the detailed modelling that we present below we shall attempt
to fit all the available kinematic constraints through a star-disc
encounter  (i.e we assume that the system is either highly eccentric or marginally unbound. For justification of this assumption, see Section \ref{subsec:init_cond}), but note that, since the errors on the proper motion data in \citet{2012ARep...56..686B}
are large, we shall place more emphasis in fitting the radial
 component of the stars' relative space motions on the plane of the sky.

  Interest in the RW Aurigae system has recently been revived by
the photometric observations of \citet{2013AJ....146..112R}. Their optical
monitoring detected a 2 magn dimming event lasting $180$ d which
the authors intriguingly interpreted in terms of occultation of RW Aur A
by a portion of the tidal arm. We shall investigate whether our best
model for the encounter involves a geometry that would permit such
an event.

In this paper we use new hydrodynamical models and synthetic observations
to fit many of the parameters of RW Aur. We aim to understand the geometry
of the binary system, the mechanism/geometry responsible for its morphological
appearance, other observed properties (such as molecular line profiles) and the recent 
dimming event found by \citet{2013AJ....146..112R}.  The rest of this paper proceeds as follows. 
In Section \ref{sec:obs} we describe the
range of observational constraints of the RW Aur system which we shall use
to constrain our models. In Section \ref{sec:num_met} we outline the codes used in our
investigation (the \textsc{phantom}\footnote{\url{http://users.monash.edu.au/~dprice/phantom}} smooth particle hydrodynamics code 
which is used for the hydrodynamical simulations
and the \textsc{torus}\footnote{\url{http://www.astro.ex.ac.uk/people/th2/torus_html/homepage.html}} Monte Carlo radiation transport code which is used to
produce synthetic observations from the results of the hydrodynamic models) and also the initial
conditions and parameter space surveyed. In Section \ref{sec:results_hydro} 
we present
and discuss our best-fitting model solution and the degree to which
various system parameters are constrained by the observations.
In Section \ref{synthobs} we present the detailed line profiles and continuum emission map for
our best-fitting model. Section \ref{sec:dimming} discusses a possible origin of the `dimming event' in the light curve of RW Aur, and whether it could be associated with occultation by the tidal arm. Section \ref{sec:concl} summarizes our
conclusions. 

\section{Observations of RW Aur}

\label{sec:obs}

 \subsection{The disc around star A}
 
 \label{sec:star_a}

The $^{12}$CO$(J=2-1)$ emission from close proximity to RW Aur A shows a perfect point symmetry between its redshifted and blueshifted lobes. Furthermore, the radial velocity gradient is perpendicular to the optical jet. Both of these features suggest an axisymmetric geometry, i.e. RW Aur A most likely has a rotating disc \citep{2006A&A...452..897C}.
The sky inclination of the disc was constrained using 
the ratio of the radial and tangential velocities of emission knots in its jet 
\citep{2003A&A...405L...1L}. Alternatively, the inclination was also constrained by the radial velocity of the jet itself. Combining these, the sky inclination was estimated to be in the range of $45-60^{\circ}$ \citep{2002ApJ...580..336W}.

The observed flux densities in $^{12}$CO$(J=2-1)$  and $^{12}$CO$(J=1-0)$
 has a ratio of about $4$ \citep[see fig. 4 by][]{2006A&A...452..897C}. The ratio of $4$ is the square of the frequency ratio between these two lines. This suggests that the disc is optically thick in which case the Rayleigh-Jeans limit predicts that $B_\nu = 2kT\nu^2/c^2$.
\citet[][hereafter BS93]{1993ApJ...402..280B} predicted that for an optically thick, Keplerian, vertically isothermal disc of radial temperature profile $T \propto R^{-q}$, the molecular line profile scales as $\nu^{3q-5}$ towards the wings of the line profile.
The disc line profile observed by \citet{2006A&A...452..897C} suggests that $q \sim 0.73$. The same parameter $q$ was also independently estimated as $ \sim 0.57$ from the infrared spectral energy distribution \citep{1995ApJ...439..288O}.
The BS93 model also predicted that the peak flux density of disc line profiles occurs at the velocity of the disc's outer radius:
$V_{\rm peak} = \sin {(i)} (GM/R_{\rm out} )^{0.5}$. 
The observed peak for A's  disc ($\sim 3.5$ km$/$s) then
constrains  $R_{\rm out}$  to be in the range of $50-77$ AU respectively for an inclination of $45-60^{\circ}$. \citet{2006A&A...452..897C} noted that stellar mass and outer disc radius are degenerated when considering the observed peak velocity. Therefore they argued that the better approach is to compare the shape (rather than absolute peak velocity) of observed and theoretical line profiles in BS93; they claimed $R_{\rm out}$ to be $40-57$ AU.

\citet{2006A&A...452..897C} estimated the temperature of the disc at its outer radius with two methods:\\

\noindent1) Measuring the peak flux density of the disc line profile and using the BS93 model which predicted that the peak flux density scales as $F(\lambda)  \sim \lambda ^{-2} R_{\rm out} ^{5/2} T_{\rm out}^ {3/2} M_{\rm star}  ^{-0.5} d^{-2} $;\\

\noindent2) Using the brightness temperature since the disc is
optically thick and in the Rayleigh-Jeans regime.\\

\noindent The two methods consistently gave a temperature of $60-107$\,K (the range corresponding  to different sky inclination) at the outer radius of the disc.

\citet{2006A&A...452..897C} further estimated the column density of the disc. The fact that 
the disc is optically thick in $^{12}$CO and undetected in $^{13}$CO placed lower and upper limits on the column density respectively. These  CO column density limits were
converted to total column densites using 
standard $^{12}$CO and $^{13}$CO abundances in the local interstellar medium. Assuming that the disc surface density scales as $1/R$, \citet{2006A&A...452..897C} derived
an  estimate  of the total mass of the disc of $10^{-5}-10^{-4}  M_\odot$. This is consistent with the mass estimate by \citet{1995ApJ...439..288O} 
($3 \times 10^{-4} M_\odot$), but not with the higher estimate by \citet{2005ApJ...631.1134A} ($4 \times 10^{-3} M_\odot$). 

 \subsection{The tidal arm}
 
 \label{subsec:tidal_arm}

 $^{12}$CO molecular emission from the tidal arm is entirely redshifted \citep[with the peak velocity at $3.2-3.8$ km$/$s with respect to star A,][]{2006A&A...452..897C}. The observed radial velocity is larger than the escape velocity at the corresponding distance, indicating that the motion of the tidal arm is unbound. The tidal arm is also optically thick as the observed $^{12}$CO$(J=2-1$)/ $^{12}$CO$(J=1-0)$ flux density ratio is $\sim 4$. The tidal arm extends about $ 4^" (\sim 600) $ AU at the distance of $140$ pc.

 \subsection{The molecular complex around star B}
$^{12}$CO molecular emission  around star B has a broad and asymmetric peak blueshifted by $3-6$ km$/$s with respect to star A. The asymmetric peak cannot be attributed to an undisturbed rotating disc. Given that the tidal interaction hypothesis is favoured, the molecular complex around B most likely consists of material captured during the tidal encounter. The $^{12}$CO$(J=2-1$)/ $^{12}$CO$(J=1-0)$ ratio was observed to be $\sim 5.7$ suggesting that the molecular complex around B is marginally optically thick \citep{2006A&A...452..897C}. Note, however, that the signal to noise of the $^{12}$CO$(J=1-0)$ emission line is very low, we therefore do not use this line as a constraint to discuss our model.

\begin{figure}
\hspace{-10pt}
\includegraphics[width = 9cm]{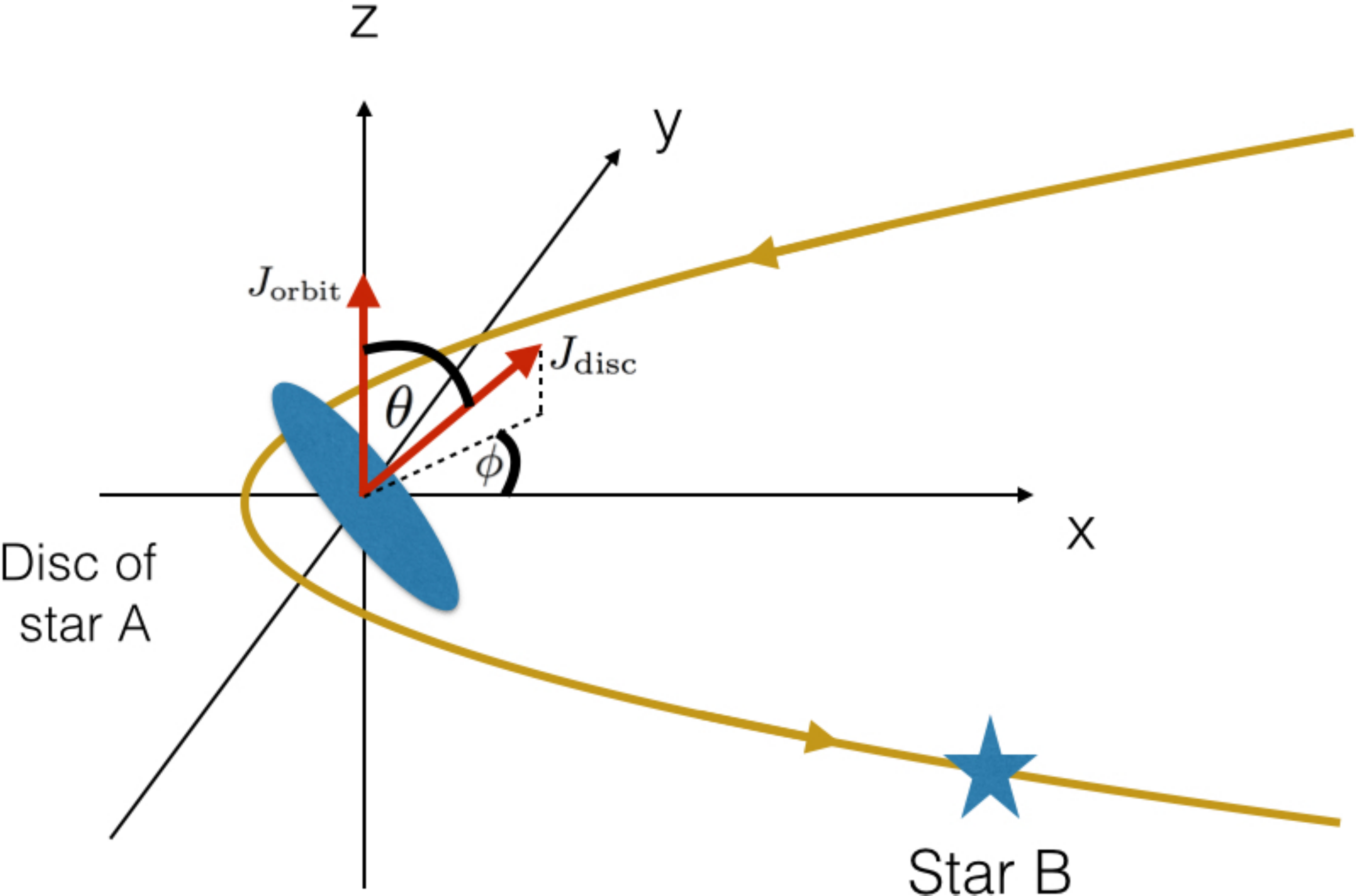}

\caption{A schematic illustrating the polar and azimuthal angle used to define the mutual inclination between the angular momenta of the disc and the orbit of the secondary.
\label{geometry}}
\end{figure}

 \subsection{Proper motion of the system}

\citet{2012ARep...56..686B} compiled the position angles of star B and the angular separations between star A and B over the past 70 yr. Linear regression analysis showed that the projected AB separation was increasing at $0.002^"/$ yr. 
$70$ yr is short compared to the Keplerian orbital period ($\sim 2000$  yr) at the observed separation ($\sim1.5^"$, or $200$ AU for $140$ pc away), 
thus justifying  a linear approximation for the velocity. 
They similarly claimed that the position angle of star B was {\it increasing}
at a mean rate of $0.02^{\circ}/$yr, which would imply that the orbital motion
of the pair was {\it anti-clockwise} as projected on to the plane of the
sky. This conclusion is however not firm given that the claimed increase
is based on a fit involving data from the 1940s whose radial component
was discrepant with the recently observed trend of increasing separation.
Moreover this fit to the tangential data is in the opposite sense to the apparent trend observed
during the 1990s (although the  uncertainties during this period were large). We therefore only constrain our model such that the sky-projected separation of star A and B is increasing at the current epoch. Future proper motion observations are needed to decide whether the motion of the two stars projected into the plane of the sky is clockwise or anti-clockwise.

\subsection{The dimming event}
\label{sec:dimm_obs}

\citet{2013AJ....146..112R} reported a long ($\sim 180$ d) and deep (2 mag) dimming of RW Aur from 2010 September until 2011 March. Typical photometric variations of CTTS (Classical T Tauri Stars) only last for days and weeks with a depth $<1$  mag. This is usually associated with circumstellar extinction or hotspots on the accretion disc that rotate into and out of the observer's view. \citet{2013AJ....146..112R} noted that there was no similar dimming event\footnote{Note that the long term light curve does show a number individual photometric data points at the `low' level recorded in the 2011 event but a well resolved eclipse light curve has not been detected hitherto.} for the system in the past $50$ yr
and proposed that the dimming  may be due to the occultation of star A by the leading edge of the tidal arm. They further estimated that the occulting body was moving with a transverse velocity of $0.86-2.58$ km$/$s (dividing the stellar radius of star A by the ingress time of the dimming). The size of the occulting body was estimated to be $\sim 0.09 - 0.27$ AU using the product between the duration of the dimming and the derived velocity. The large range in the estimates of transverse velocity and size are due to the large uncertainties on the ingress time-scale, approximately between $10$ and $30$ d.

More recently, \citet{2015IBVS.6126....1A} have again detected RW Aurigae
in the `low state' \citep[i.e. 3-4 mag less than what observed by HST in the mid-90's,][]{2001ApJ...556..265W}.

\subsection{ Summary of observational constraints}

Among the various observations, we identified the following key criteria to gauge how closely our simulations describe the observations. These features were more reliably constrained by observations; they are also easily quantifiable in simulations:\\

\noindent1) The size of the system:
\citet{2006A&A...452..897C} reported that the tidal arm extended about $4^" (\sim 600$ AU at a distance of $140$pc); the projected AB separation was $\sim 1.5^"$ ($ \sim  200$ AU). We aim to reproduce the size of the system as observed. However we argue that it is more important to fit the relative ratio between the dimensions of various structures of the system. This is because the relative ratio is independent of the error in the distance measurement  \citep[a parallax of $7.08 \pm 0.71$   mas corresponding to a  distance of $140 \pm 14 $ AU][]{1989ApJ...341..340B}.\\

\noindent2) The relative orientation of the A disc and star B.
We gauged the relative orientation using position angles (PA). For the disc, its position angle is defined as the angle between its radial velocity gradient 
and north. For star B, PA is defined as the angle between the vector joining star A and B and North. According to \citet{2006A&A...452..897C}, the PA of the disc is $\sim 40^{\circ} $; the PA for B is $\sim 105^{\circ}$ \citep[or $255^{\circ}$ if using the convention of][]{2012ARep...56..686B}.\\

\noindent3) Sky inclination of the disc:
the sky inclination of the disc was estimated to be in the range of $45-60^{\circ}$ \citep{2002ApJ...580..336W,2003A&A...405L...1L}.\\

\noindent4) Proper motion:
as reported by \citet{2012ARep...56..686B} the angular separation between star A and star B was increasing at a rate of $0.002\arcsec/$ yr.\\

\noindent5) The line profiles:
the radial velocity of the peak, the peak flux density and the optical depth.

\section{Numerical Method}
\label{sec:num_met}
\subsection{Hydrodynamics models}
We model the stellar fly-by using the smooth particle hydrodynamics (SPH) code 
\textsc{phantom} \citep[see e.g.][]{2010MNRAS.405.1212L, 2010MNRAS.406.1659P,2012JCoPh.231..759P,2013MNRAS.433.2142F,2013MNRAS.434.1946N}.
We assume that self-gravity in the disc is negligible and that the gravitational
potential is dominated by the stars.
We employ a locally isothermal equation of state
and adopt standard artificial viscosity parameters ($\alpha= 0.1; \beta = 2;
\sigma = 0.1$) according to the presciption of \citet{1997JCoPh.136...41M}.

\begin{figure}
\centering
\includegraphics[width = \columnwidth]{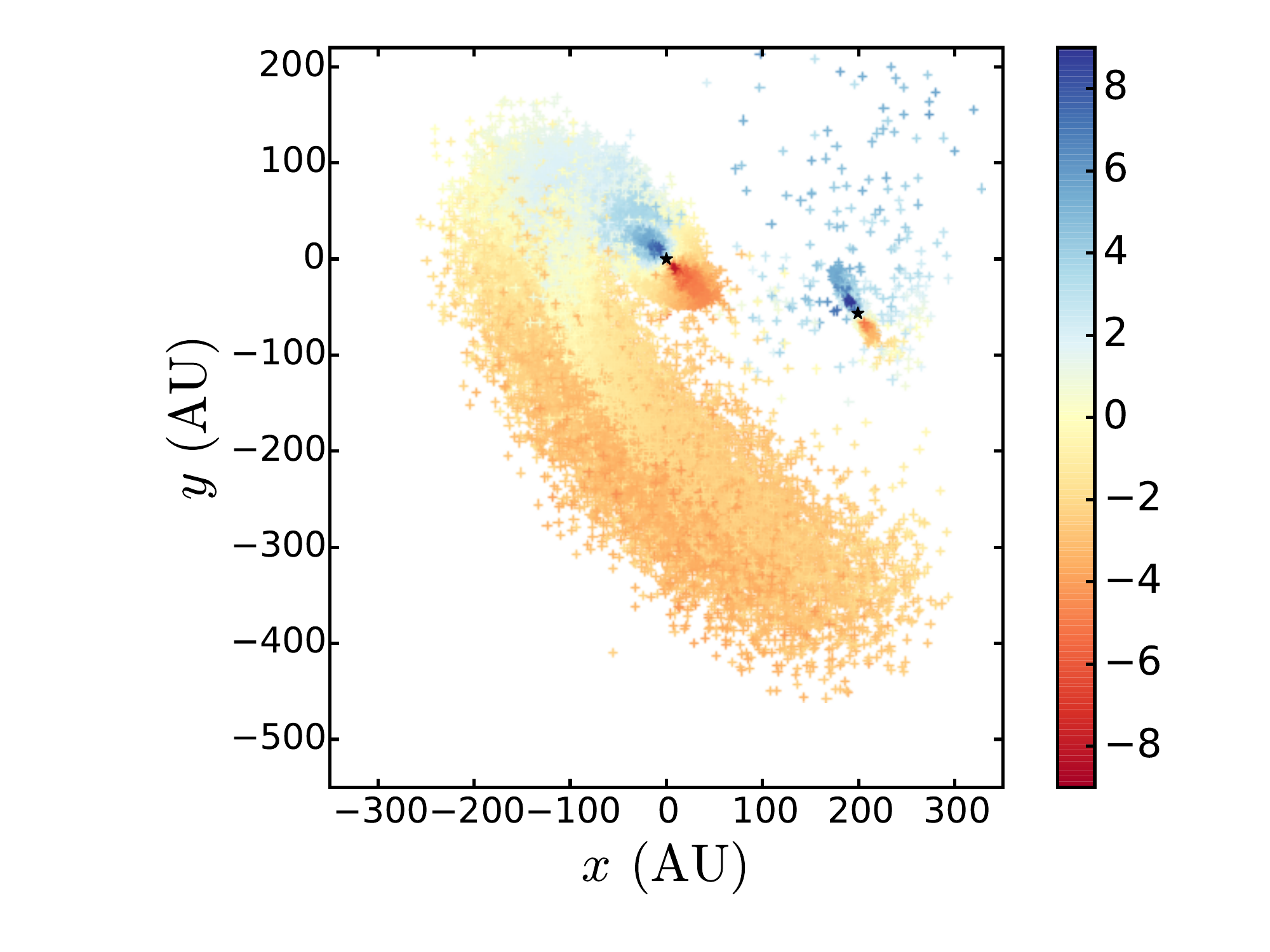}
\caption{The gas morphology of the best-fitting model in the frame of reference comoving with star A. The $x$-$y$ plane indicates the plane of the sky. Crosses indicate SPH particles, which were selected randomly every other 10 particles. The black stars show the position of Star A and B. Colour coding is linear in radial velocity (km/s). Note, for direct comparison with Fig \ref{fig:cabrit}(c), the heliocentric velocity \citep[$15.87\pm0.55$\,km/s,][]{2006A&A...452..897C} of the star A should be added.
\label{radial}}
\end{figure}

\subsection{Synthetic observations}
\label{torussec}
We use the radiation transport and hydrodynamics code 
\textsc{torus} to compute synthetic observations in this paper.
\textsc{torus} is predominantly a Monte Carlo radiation transport
code used for calculating line and continuum transport \citep[e.g.][]{2000MNRAS.315..722H, 2006MNRAS.370..580K} 
but also features photoionization \citep{2012MNRAS.426..203H}, non-LTE molecular line transfer \citep{2010MNRAS.407..986R} and hydrodynamics \citep{2012MNRAS.420..562H}.

We map the result of the SPH simulation on to the \textsc{torus} grid using the technique described in \cite{2010MNRAS.403.1143A}. We do not explicitly derive the temperature using radiative transfer, but rather assume that the temperature distribution in the model is a function of cylindrical
distance from the primary (see below;  equation \ref{temperature}), as
in \cite{2006A&A...452..897C}.
For molecular line transfer calculations we solve the equations of non-LTE statistical equilibrium and produce synthetic data cubes in the manner described in detail by \citet{2010MNRAS.407..986R}. 

Since we are comparing with the observations from \citet{2006A&A...452..897C} we consider the $^{12}$CO$(J=2-1)$ and $(J=1-0)$ lines and assume the same parameters, i.e. the  $^{12}$CO abundance relative to hydrogen is $8\times 10^{-5}$ and the system is at a distance of 140  pc from the observer. We assume a small micro turbulence value of 0.03   km$/$s.
The data cubes comprise $401 \times 401$ pixels and $36$ velocity channels spanning $-7.19$ to $7.63$\,km$/$s. This velocity range and number of channels is chosen to match the IRAM Plateau de Bure Interferometer, the 
instrument used for the observations in \citet{2006A&A...452..897C} against which we compare our models. The beam size is chosen to match that of the IRAM Plateau de Bure Interferometer ($2.42\arcsec \times 1.55\arcsec$ and $0.89\arcsec \times 0.58\arcsec$ for the $J=1-0$ and $J=2-1$ transitions respectively). We convolve the resulting data cubes with a 2-D Gaussian with a width corresponding to the semiminor axis of the two beams using \textsc{aconvolve} from \textsc{ciao} v 4.5 \citep{2006SPIE.6270E..60FB}. Finally, the line profiles of the synthetic observations are obtained by integrating over ellipses that match the ones used by \citet{2006A&A...452..897C}.

We also compute a 1.3\,mm continuum image using Monte Carlo radiative transfer, algorithmic details of which are given in e.g. \cite{2000MNRAS.315..722H} and \cite{2004MNRAS.350..565H}. We discuss the assumed dust distribution in Section \ref{dust}.

\begin{figure}
\centering
\includegraphics[width = \columnwidth]{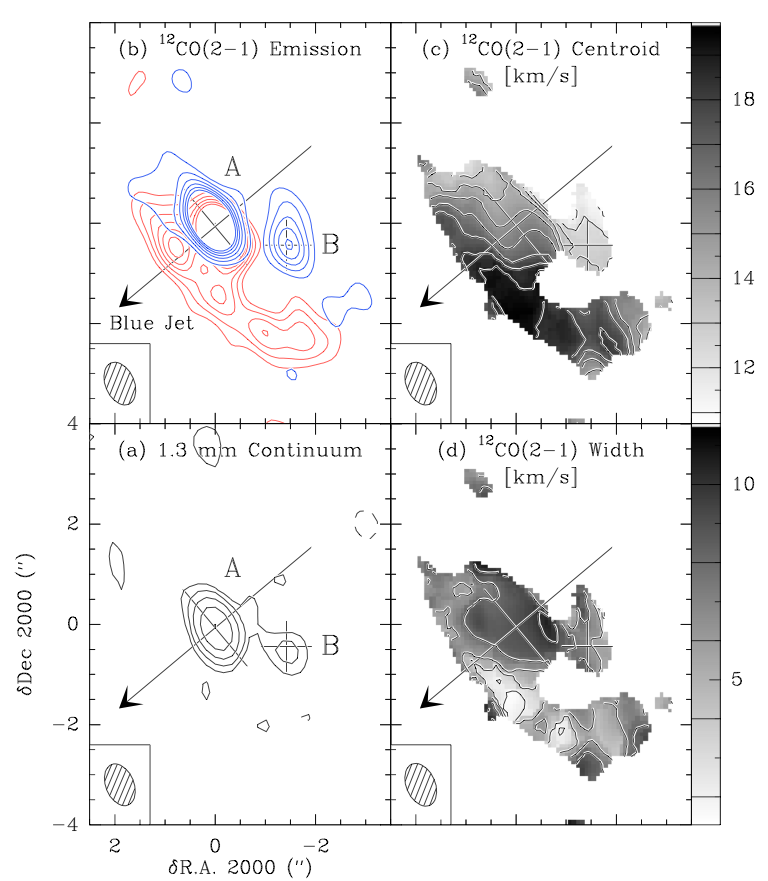}
\caption{Fig. 1 from \citet{2006A&A...452..897C}, showing the interferometric data of PdBI. From their caption: ``{\bf a)} $1.3$\,mm continuum map of the RW Aur system. Contour levels are $3$, $6$, $12$, $24$, $48\sigma$ ($1\sigma = 0.47$\,mJy/beam). Crosses indicate the peak positions, which closely correspond to RW Aur A and B. The large arrow shows the direction of the blueshifted jet. {\bf b)} $^{12}$CO$(J = 2-1)$ emission blueshifted and redshifted with respect to the RW Aur A systemic velocity. Contour spacing is $3\sigma$ ($0.12$\,Jy km s$^{-1}$ ), truncated to $18\sigma$ for clarity. {\bf c) and d)} Heliocentric centroid velocity and line width (i.e. first and second order moments) of the $^{12}$CO$(J = 2-1)$ line.'' } 
\label{fig:cabrit}
\end{figure}

\subsection{Initial conditions}

\label{subsec:init_cond}

We model the encounter geometry shown in Fig \ref{geometry}. in which
a spherical polar system (with the z-axis along the normal to the plane
of the secondary's orbit, and the x-axis pointing towards the apocentre) is used to define the coordinates ($\theta,\phi$)
of the disc normal (where the separation vector of
the two stars at pericentre is located at $\phi= \pi$).
We populate particles in a Keplerian disc around the primary according
to a prescribed surface density law (see below) and with a Gaussian
distribution normal to the disc plane which is the hydrostatic
equilibrium solution set by the imposed local temperature (see below). 
For reasons of computational economy we set an inner radius of $5.9$AU
and do not continue to integrate particles that stray within this
radius. 

\begin{figure*}
\begin{center}
\includegraphics[width=.9\columnwidth]{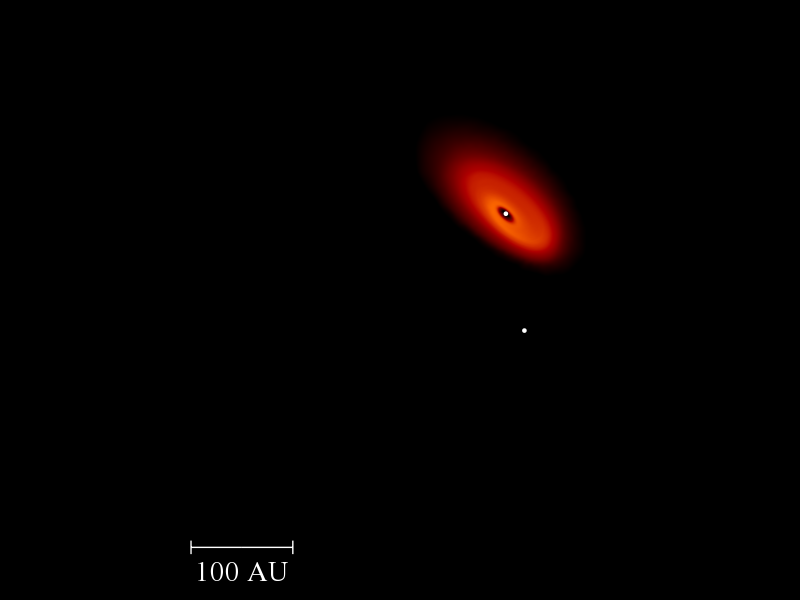}
\includegraphics[width=.9\columnwidth]{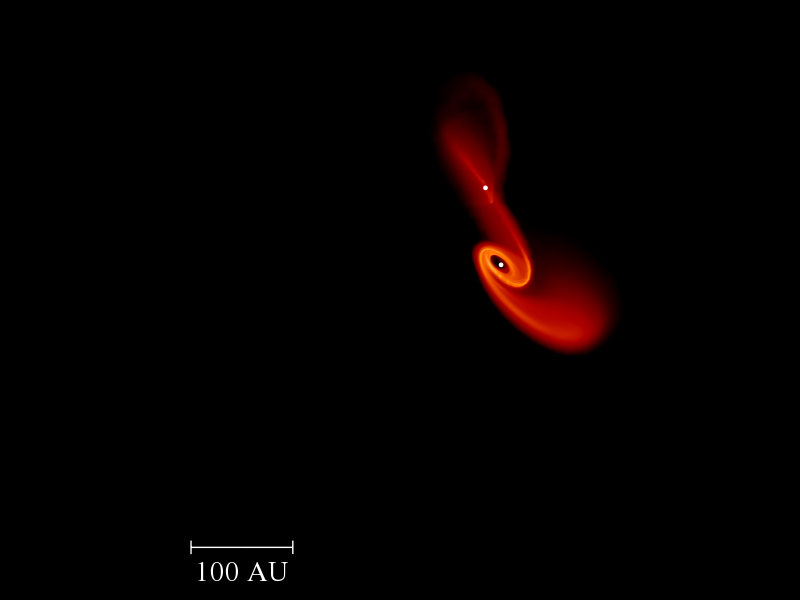}\\
\includegraphics[width=.9\columnwidth]{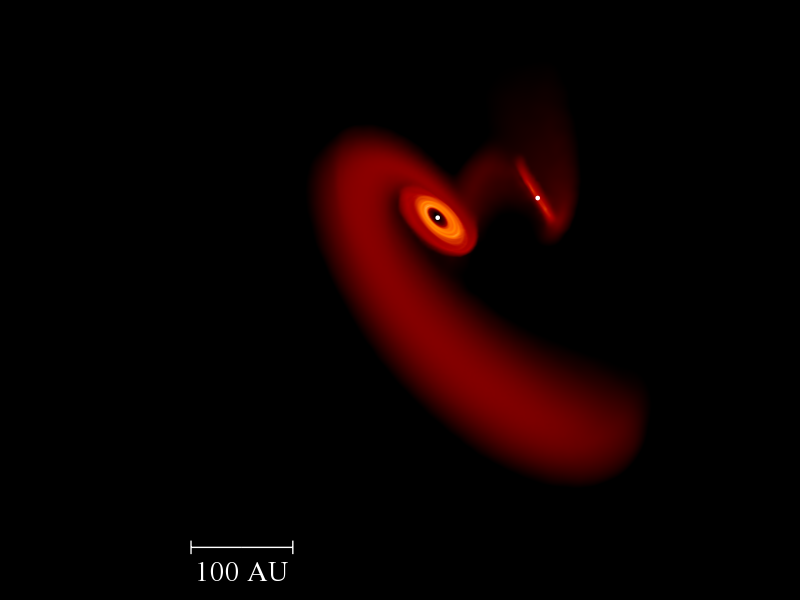}
\includegraphics[width=.9\columnwidth]{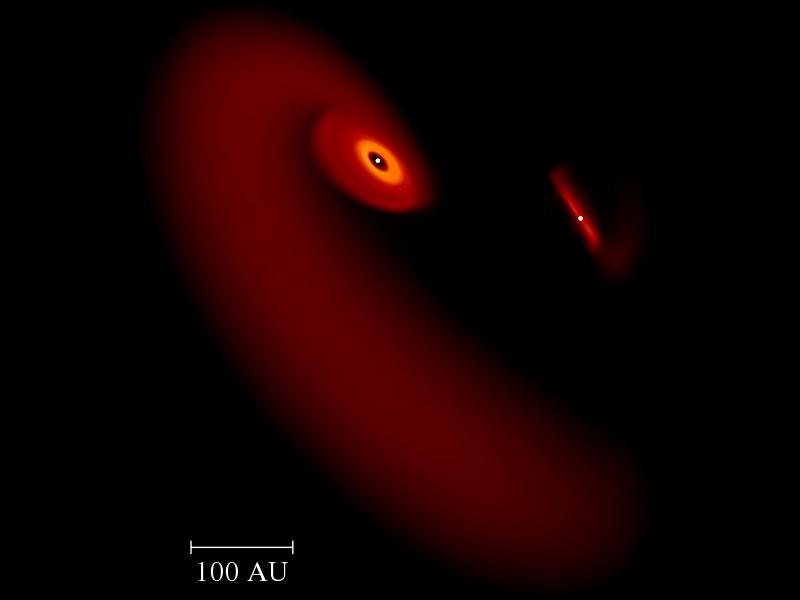}
\end{center}
\caption{Evolution of the tidal encounter in the plane of the sky, shown with four different snapshots. The bottom right panel shows the structure as seen today. From top left to right bottom the snapshots are taken at $-589$, $-398$, $-207$ and $0$ yr from today, respectively. The surface density is logarithmically scaled, and was chosen such that the final snapshot resembles the observations by \citet{2006A&A...452..897C}. Note that a stream of material linking star A and star B is present, even though it does not appear in the bottom right panel, due to the choice of surface density threshold (compare this final snapshot with the right-hand panels of Fig. \ref{fig:occultation}).   }
\label{fig:sph}
\end{figure*}

We argue that the presence of a single well-defined tidal feature in the
observations suggests that the disc of RW Aur A has been subject to
either a one-off encounter or to an encounter that is well separated
in time from any previous interactions. Thus we consider it likely that
the orbit is either of high eccentricity or unbound. It is unlikely 
that the encounter is {\it highly} hyperbolic since this would imply a
relative velocity at infinity that exceeds typical stellar velocity dispersions
in star forming regions. We thus expect that the orbit - whether mildly
bound or mildly unbound - is likely to be close to parabolic and hence
model this situation here.This is consistent with the low relative radial velocity between the two stars reported by \citet{2004ApJ...616..998W}. The masses of RW Aur A and B are set to
$1.4M_\odot$ and $0.9M_\odot$ respectively \citep{1997ApJ...490..353G,2001A&A...376..982W}. 

 We set the surface density profile to $\Sigma \propto 1/R$ and adopt a power
law temperature profile as a function of cylindrical distance from
the primary parametrized according to

\begin{equation}
T =T_0 (R/R_0)^{-q}
\label{temperature}
\end{equation} 

 The radial index, $q$ has been estimated to be in the range
$0.5-0.73$ by various authors \citep[$q = 0.57$ and $q = 0.73$][respectively]{1995ApJ...439..288O,2006A&A...452..897C}. We 
adopt a temperature around $60$\,K at $R=57$\,AU in line with the estimate
of \citet{2006A&A...452..897C} for an inclination of the disc $i=60^\circ$, and vary the value of $q$ (which  affects 
the pressure scale length of the outer disc and has a modest effect
on the morphology of the tidal tails produced).

  The simulation is initiated when the stellar separation is
 $\geq 10$ pericentre radii in order to ensure that the disc is not
violently perturbed by the sudden introduction of the companion
in its vicinity.

\subsection{Constraining other parameters}
After setting up our model with the plausible initial conditions discussed in the previous section, in principle one could constrain the various other orbital parameters by running the hydrodynamic simulation and synthetic observations in conjunction with a parameter study tool (e.g. Markov chain Monte Carlo). However, such a parameter study is unfeasible in our case: conventionally, $10^6$ particles are needed for sufficient resolution in such a hydrodynamic simulation. In conjunction with the simulated observations, this would take several days of computation per model. Moreover there are at least 10 remaining parameters in the model and so the time required for such an approach would be prohibitive.

Instead we adopted the following strategy. We initially ran our hydrodynamic code with $10^4$ test particles on a coarse grid in the 10-D parameter space. We then focused on the region that is most consistent with various observational constraints. Parameters were then tuned until simulated values converged to observed values within respective uncertainties. Finally we repeated the same model  with  $10^6$ test particles and appropriate scalings (e.g. viscosity scales with number of particles) to reveal small-scale details of the system.

We remind the reader that although we refer to this final model as the best-fitting one, in reality this is only the best-fitting model among the few dozen models we tested. Given the uncertainties in the observations
(Section \ref{sec:obs}) and the various shortcomings of our simulations,
the statistical errors introduced from our sampling strategy are likely
to be outweighed by other factors (see discussion in Section 5).
We nevertheless argue that the good level of agreement
between simulations and observations in the models performed is already sufficient to demonstrate that the tidal encounter hypothesis is indeed consistent with the existing observations.

\section{Results of hydrodynamics models}

\label{sec:results_hydro}
 
In this section, we first consider the parameters of the encounter that affect 
the gross properties of the flow
(i.e. the morphology and kinematics of the tidal tail and disc). 
 It is sufficient for such an analysis to be conducted with the hydrodynamic code alone in the limit
of optically thin emission. We fix the stellar masses and eccentricity  of the orbit
for all simulations and for each simulation we adopt given values of
$\theta$ and $\phi$, of pericentre distance, disc outer radius, surface density and temperature normalization, and temperature power-law index. In Section 5, we consider whether a plausible initial disc model can then
reproduce line fluxes and line profiles by using synthetic observations for
a range of output times and viewing angles (constrained so as to give
the correct projected separation of the two stars on the sky). The viewing angle is parameterised
in terms of the direction of the line-of-sight vector identified by the polar coordinates $\theta_{\rm obs}$ and $\phi_{\rm obs}$ in the coordinate frame shown in Fig. \ref{geometry}.

\subsection{Sensitivity of results to parameter choice}

\begin{figure*}
\begin{center}
\includegraphics[width=\columnwidth]{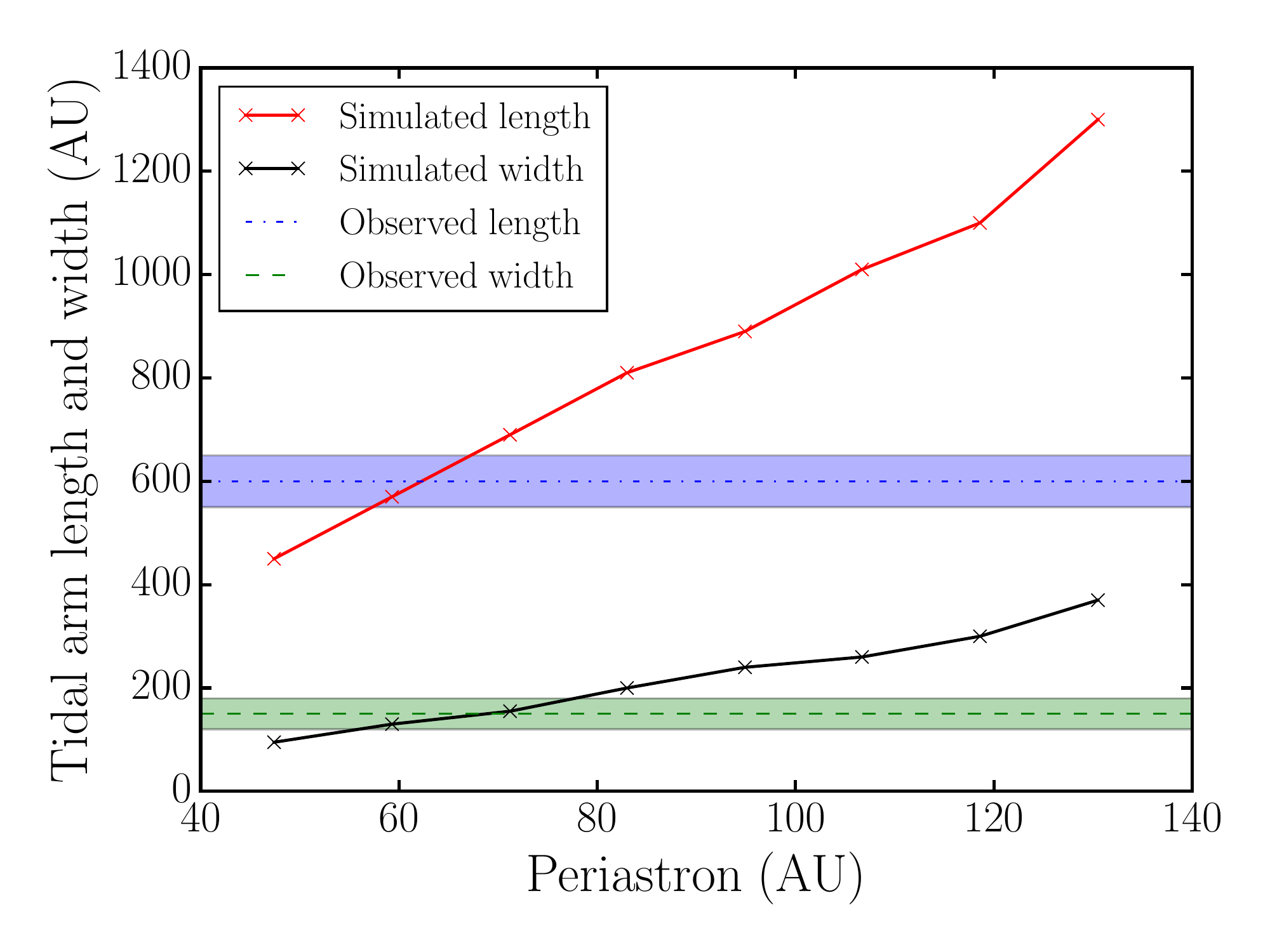}
\includegraphics[width=\columnwidth]{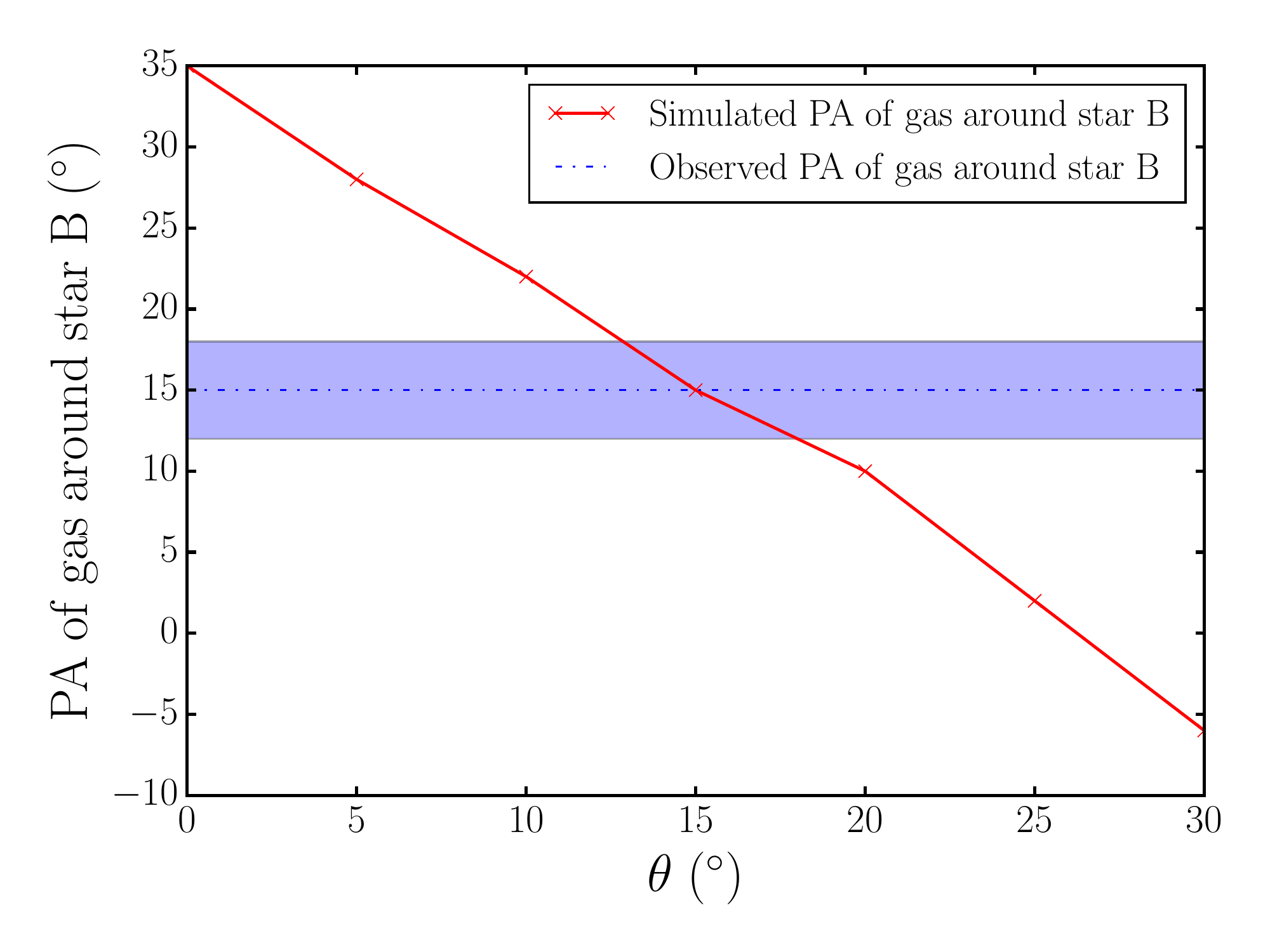}\\
\includegraphics[width=\columnwidth]{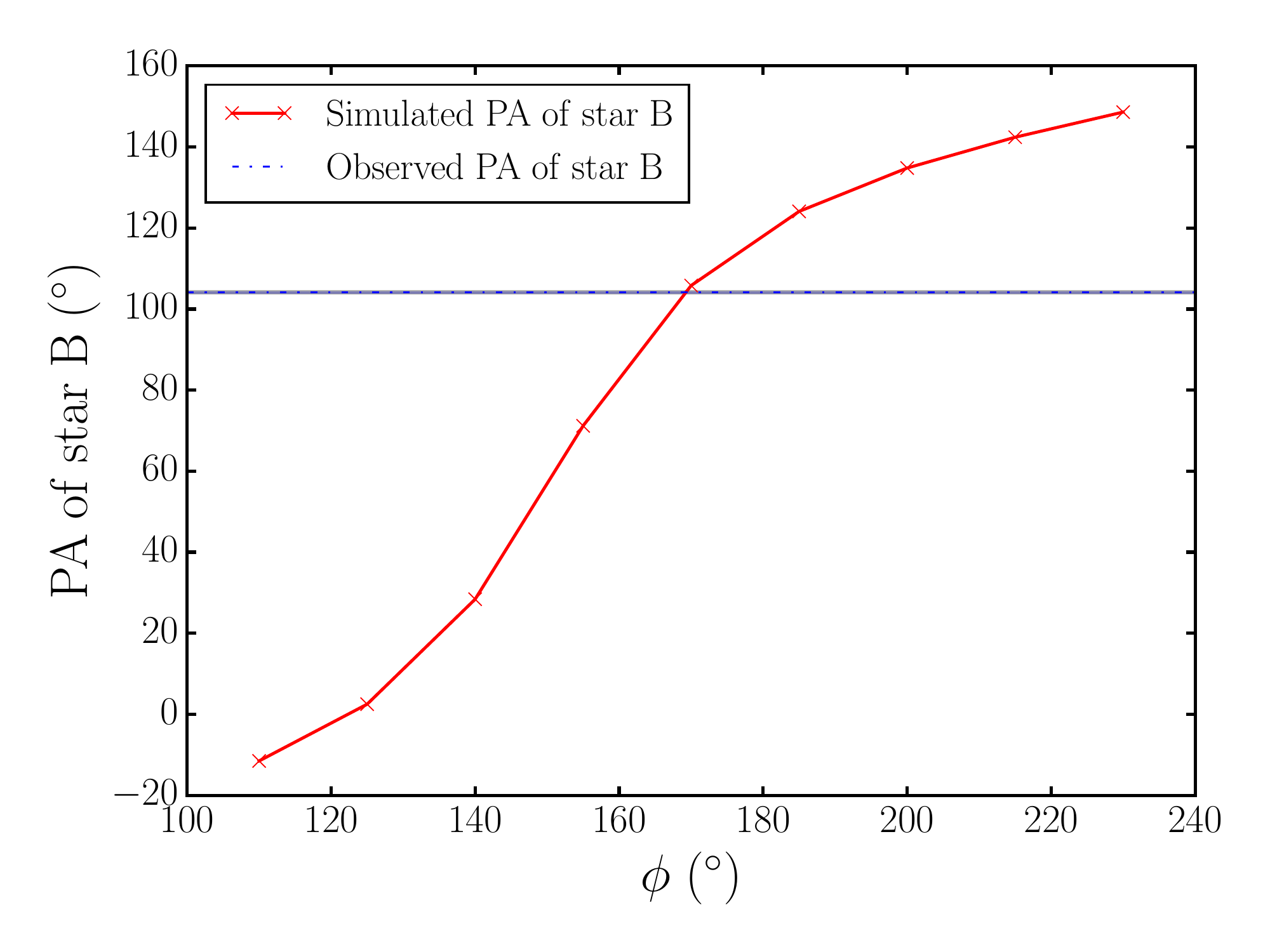}
\includegraphics[width=\columnwidth]{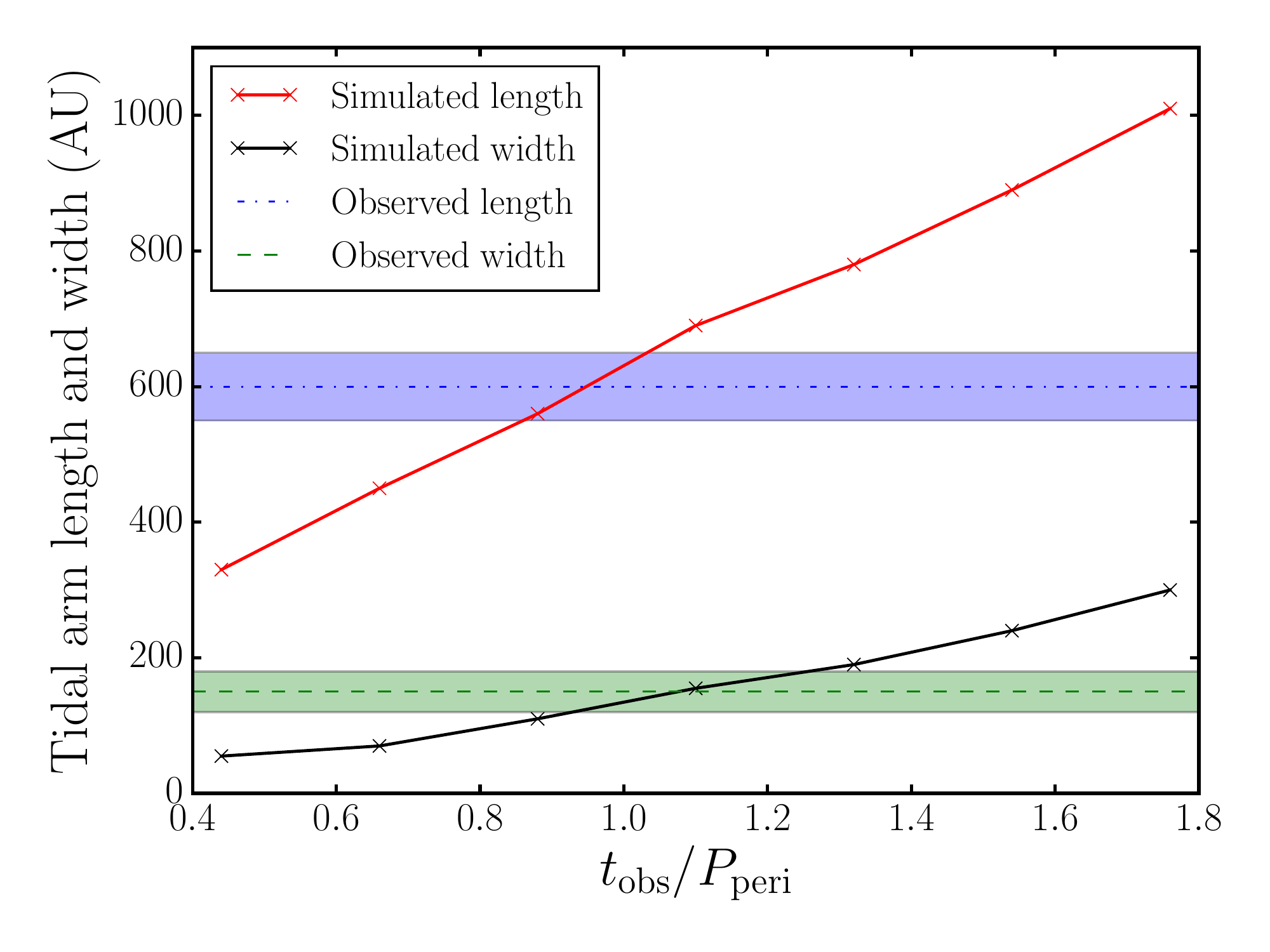}\\
\includegraphics[width=\columnwidth]{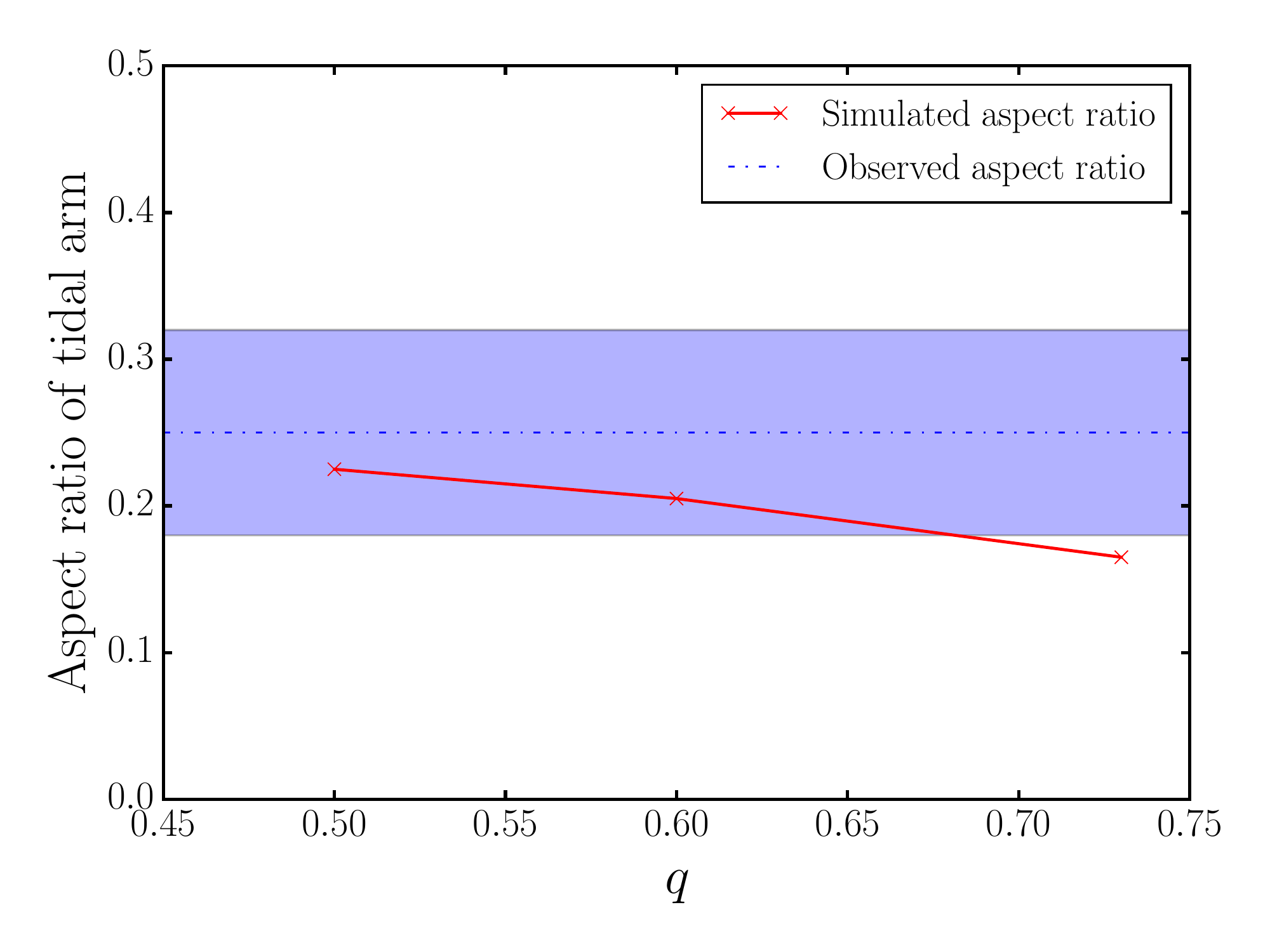}
\end{center}
\caption{In each of the diagrams, one parameter of the model is varied at a time and plotted against the observational feature that is most dependent on it. Shaded areas show observed values within $1\sigma$. Comparison with observed values helps to constrain the parameter being varied (see Table \ref{tab:comparison} for the best-fitting model). Top left: the pericentre against the length and width of tidal tail; top right: the polar angle $\theta$ between the orbital and disc angular momenta against the positional angle of the captured material on star B; centre left: the azimuthal angle $\phi$ between the orbital and disc angular momenta against the positional angle of star B; centre right: the time of observation against the length and width of tidal tail; bottom: the power index q of radial temperature profile against the aspect ratio (width/length) of the tidal tail.}
\label{fig:1_D_plots}
\end{figure*}

To determine which model is most consistent with observations, we vary each parameter around the initial estimates obtained from the pilot simulations. We then identify the system observables that are most dependent on each parameter in order to constrain that parameter.

\begin{table}
\centering
\caption{Quantitative comparison of best-fitting model outputs and observed values.
$^{a}$ From \citet{2006A&A...452..897C} assuming $d=140$ pc where necessary. $^{b}$ \citet{2003A&A...405L...1L}, \citet{2002ApJ...580..336W}. $^{c}$ \citet{2012ARep...56..686B}.}
\begin{tabular}{lll}
\hline
\hline
{\rm Observed features}  & {\rm Observed values} &  {\rm Best fit simulation }  \\
& & values\\
\hline
 {\rm Length of tidal arm }  & $ 600 $\,AU $(4\arcsec)^{a} $ & $ 650\, {\rm AU} $ \\
{\rm AB separation on sky}  & $  200$\,AU $(1.5\arcsec)^{a} $ &$ 200\, {\rm AU} $ \\
{\rm PA of disc} $(^\circ)$& $40^{a}$ &$ 47.2 $ \\
{\rm PA of AB} $(^\circ)$& $105^{a}$  &$ 105.9 $ \\
{\rm Inclination of disc} $(^\circ)$&  $45-60^{b}$ & 64.0  \\
 {\rm Rate of change of } &$ 0.002\arcsec {{\rm yr}^{-1}}^{c} $ & $0.003\arcsec {\rm yr}^{-1}$ \\
\hline
\end{tabular} 
\end{table} 

The choice of periastron separation ($70$ AU) was mainly decided by fitting observed length and width of the tidal tail both of which increase with pericentre separation (see Fig. \ref{fig:1_D_plots}a). The length and width of the tidal tail is also dependent on time of observation. However this degeneracy can be broken, since the choice of periastron separation must also give a consistent radial velocity of material in the tidal tail which is
similar to the escape velocity at periastron (see Fig. \ref{fig:line_profiles}). Moreover, since the disruptiveness
of the encounter  depends on
the ratio of the periastron separation to the outer disc radius, the
observed scale of the residual disc constrains the initial disc radius 
($60$ AU).

\begin{table*}
\caption{Parameter choices that best fit the observations and the effect of
parameter variation on system properties.}
\begin{tabular}{rll}
\hline
\hline
{\rm Parameter varied}& {\rm Best fit value}  & ${\rm Effect}$ \\
\hline
{\rm Initial disc radius}& $ 60\, {\rm AU}$ & {\rm Post-encounter disc size and tidal tail morphology} \\
{\rm Periastron separation} & $70\, {\rm AU} $ & {\rm Morphology and kinematics of tidal tail}   \\
{\rm Mutual inclination of disc and binary planes} & $  \theta = 18^\circ$, $ \phi = 170^\circ $  & $\theta$ {\rm determines the orientation of tidal tail and} \\
 & & {\rm the captured material on star B;} \\
 & & $\phi$ {\rm determines the velocity and positional angle of star B} \\
{\rm Viewing angle} & $ \theta_{\rm obs} = 76^\circ, \phi_{\rm obs} = 114^\circ $ &  {\rm Morphology in plane of sky}  \\ 
{\rm Time of observation} &  $  1.1\, P_{\rm peri} $ after pericentre & {\rm Kinematic signatures and ratio of AB separation to }  \\
& & extent of tidal arm\\
 {\rm Radial Temperature profile}& $ T = 230\,{\rm K} \left(\frac{R}{{\rm AU}}\right)^{-0.5}$  &{\rm Power law index determines aspect ratio of tidal tail}  \\
 {\rm Initial Disc Mass}& $0.00167M_\odot $  &{\rm Optical depth of different regions of the system}  \\
\hline
\end{tabular} 
\label{tab:comparison} 
\end{table*}

Another important quantity that could be used to constrain the orbital parameters of the flyby is the disc outer radius, after truncation. Since \citet{1993MNRAS.261..190C} other groups have worked on the angular momentum and energy transfer mechanisms occurring with a star-disc encounter \citep[e.g.][]{2001Icar..153..416K,2015MNRAS.446.2010M}. Recently, \citet{2014A&A...565A.130B} focused on the sizes of protoplanetary discs after such encounters using numerical simulations, where the discs are described by test particles (hydro effects are neglected).  They model the encounter via prograde coplanar parabolic orbits. They obtain the following prescription for the outer radius of the truncated disc: $R_{\rm out} = 0.28 R_{\rm peri} m_{12}^{-0.32}$, where $m_{12} = M_2 / M_1$.  However, note that they did not explore the effects of the misalignment between the disc and the plane of the parabola. Recent results by \citet{2014arXiv1412.7741L}  analysed how the tidal truncation radius estimates are affected by the misalignment between a disc and an external binary \citep{1994ApJ...421..651A}. Their results indicate that the external torque decreases quite rapidly with inclination, leading to a larger circumstellar disc. Similarly, we expect the same kind of behaviour for a flyby interaction, even though in this case resonances do not come into play. Finally, the analytic prescription by \citet{2014A&A...565A.130B} strongly depends on the definition of the outer radius. They define it as the point where the surface density profile presents a maximum in its radial gradient. However nothing prevents an observer from detecting regions of the disc outside this radius, where the material will be less bound and in more elliptical orbits \citep[see again][]{2014A&A...565A.130B}. To conclude, the outer radius is not the best parameter to be used to constrain {\it a priori} the orbit of the flyby. We will check {\it a posteriori} that what we obtain is in reasonable agreement with the observations (see Section \ref{synthobs}).

The mutual inclination of the disc and the binary orbit ($\theta$)
is the main determinant of the morphology of the system. While the azimuthal angle $\phi$ determines the relative position and separation between star A and star B, the polar angle $\theta$ determines the size and orientation of the tidal arm. The tidal arm is most prominent and most massive for a prograde coplanar tidal encounter. As $\theta$  increases, the tidal arm decreases in size and mass; it also becomes warped with respect to the original disc plane. If $ \theta$  is larger than $90^{\circ}$, no extensive tidal arm is produced.

We therefore {\it require} a prograde encounter. We can obtain a clue about
the likely orientation of the disc of A on the sky from the observation
of its jet which is directed to the SE. Since the jet must be visible on the
observer's side of the disc, it suggests that the NW side of the disc is
nearer to the observer. The fact that the NE lobe of A's disc is
blueshifted then suggests that the rotation of the disc is clock-wise as
seen from the earth. In the successful (rather close to coplanar
prograde) model, we have found a viewing angle that well reproduces
the observed blueshift of the material centred on B. However, the predicted
proper motion in the plane of the sky is  then {\it
clock-wise} in contrast to that claimed by \citet{2012ARep...56..686B}.
We have discussed above how it is desirable to better constrain
the motion in the plane of the sky through future proper motion
observations.

The power index $q$ of the adopted radial temperature profile determines the rate at which temperature falls as a function of radius. A smaller value of $q$ results in a hotter and thus thicker outer region of the undisrupted disc. Subsequently, a thicker outer disc produces a thicker tidal tail after the tidal interaction. Therefore, the value of $q$ is constrained via the observed ratio between the width and length of the tidal arm.

\subsection{The best fitting model} 

\begin{figure}
\includegraphics[width = \columnwidth]{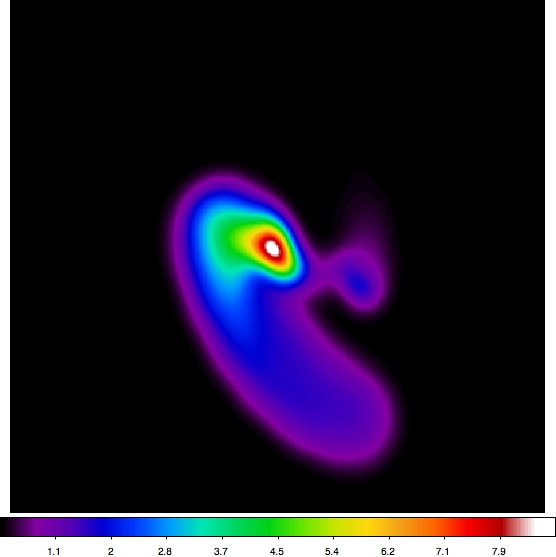}
\caption{The integrated  $^{12}$CO$(J=2-1)$ emission map of RW Aur after correcting for beam dilution. Colour coding is linear in Jy beam$^{-1}$ km/s.}
\label{1to0}
\end{figure}

Fig. \ref{radial} shows the gas morphology of the best-fitting model, the parameters of which are listed in Table \ref{tab:comparison} \citep[compare Fig. \ref{radial} with Fig. \ref{fig:cabrit}, where we portray Fig. $1$ by][]{2006A&A...452..897C}. The gas is rendered as individual SPH 
particles which are colour coded according to line of sight velocity. Fig. \ref{fig:sph} shows the evolution of the fly-by with rendered plots. 

This model successfully reproduced the following observed features of the system:

\begin{itemize}
\item{The northeast lobe of the disc is blueshifted with radial velocity of $ \sim -3$\,km$/$s with respect to Star A  while the southwest lobe of the disc is redshifted with radial velocity of $\sim 3 $\,km$/$s.}

 \item{The tidal arm has the observed spiral shape with the correct handedness. It is entirely redshifted with a radial velocity of $\sim 3$\,km$/$s 
which is larger than the escape velocity at the corresponding distance. The tidal arm is thus unbound, as observed.}

\item{The molecular complex on B is blueshifted with respect to star A. The shape of the captured  material around B is similar to that observed with its
major axis being  tilted slightly to the east of north. 
This good  agreement suggests that the gas observed around B may  indeed be 
material captured from A's disc (even if B also originally possessed
a small scale disc of its own).}
\end{itemize}

Comparing the observation and simulation more quantitatively, Table 2 lists the various quantifiable morphological features. Our simulations matches the observations in nearly all respects. The strikingly good agreement demonstrates that the system's morphology is most likely associated with a tidal interaction.

\begin{figure*}
\begin{center}
\includegraphics[width=\columnwidth]{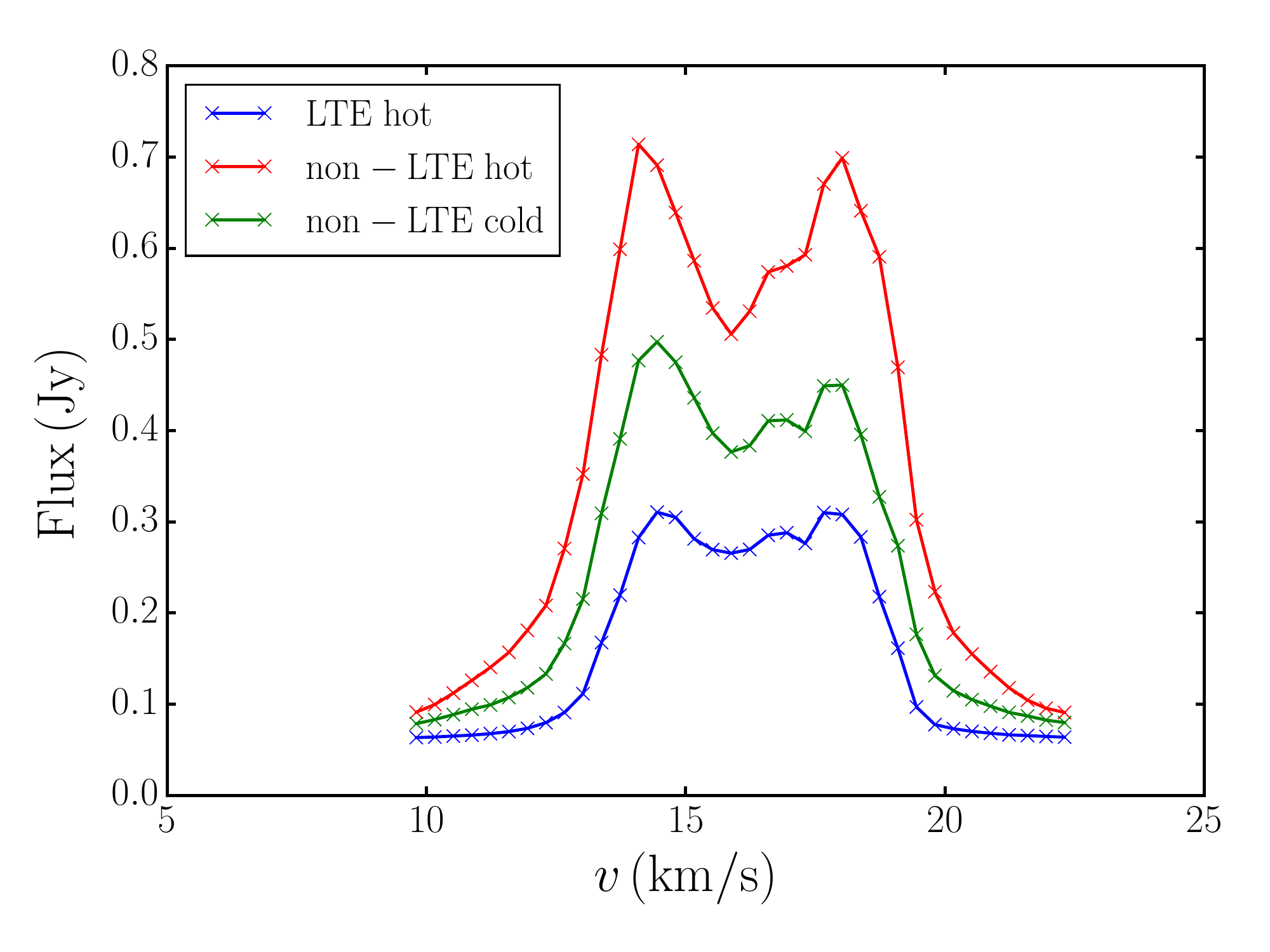}
\includegraphics[width=\columnwidth]{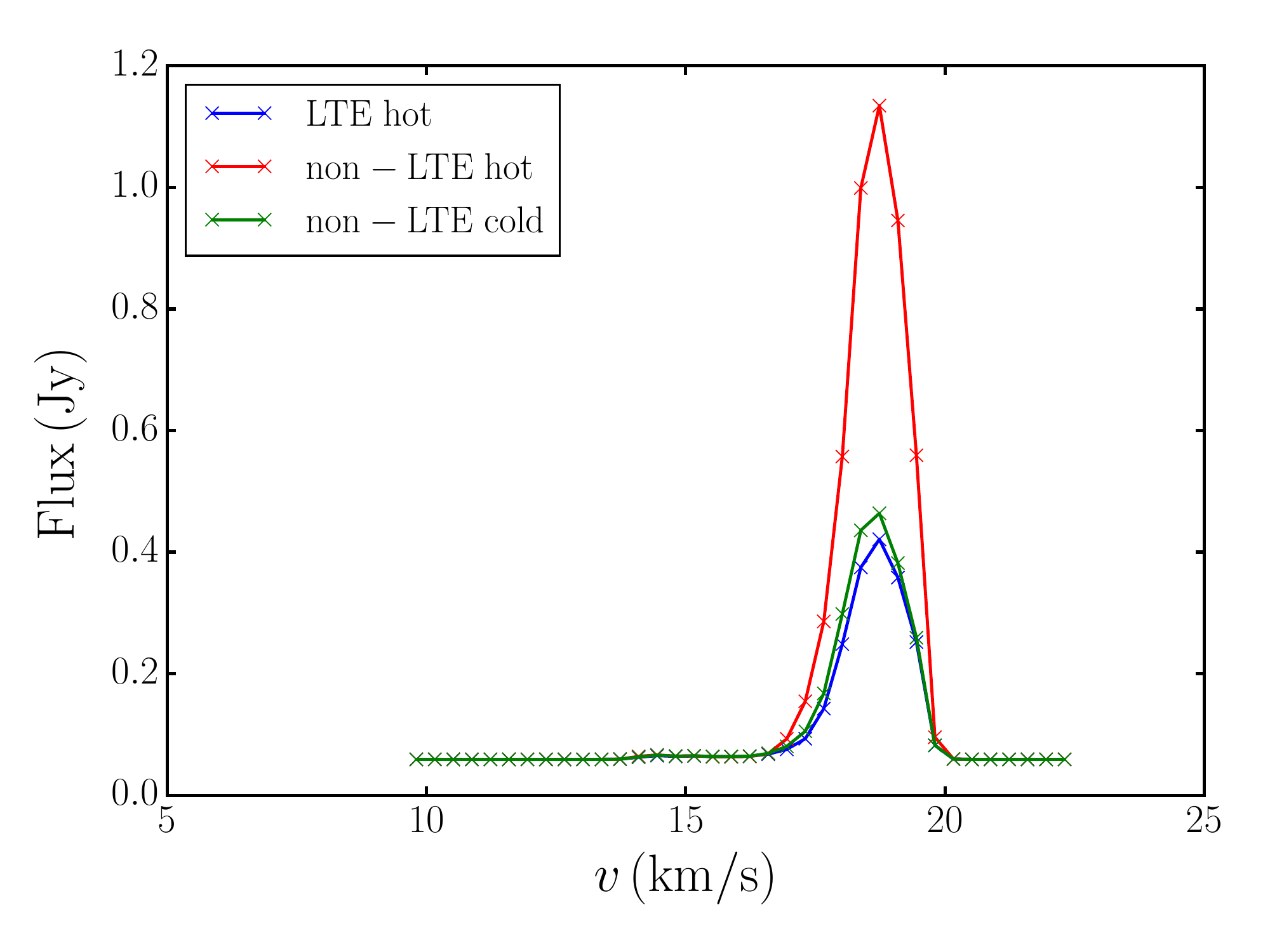}\\
\includegraphics[width=\columnwidth]{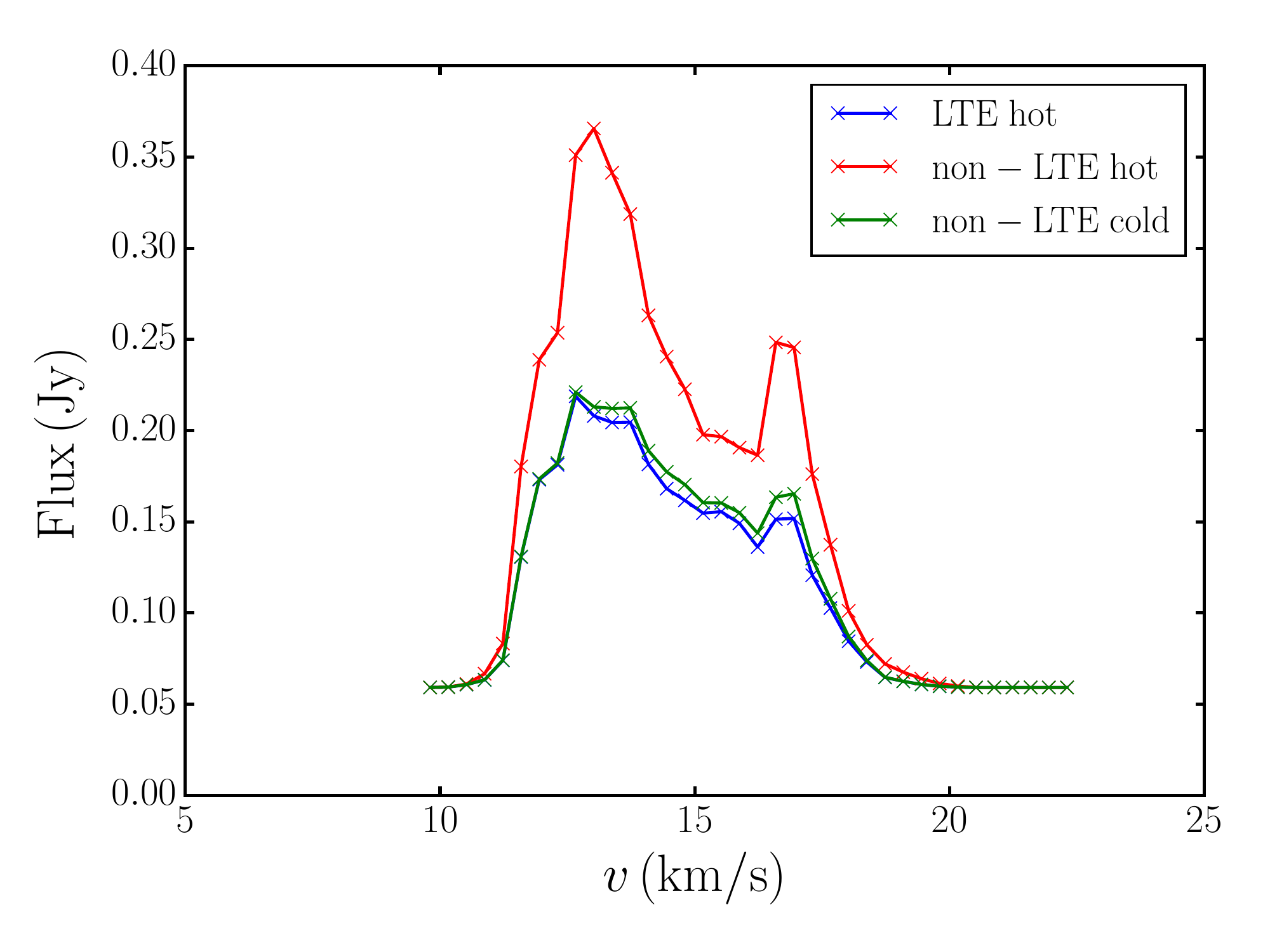}
\end{center}
\caption{The integrated line profiles of the $^{12}$CO$(J=1-0)$ emission line of the A disc (top left panel), tidal arm (top right panel) and molecular complex (bottom panel). The flux has been multiplied by a factor $4$, in order to compare these synthetic profiles with the observed ones in \citet{2006A&A...452..897C}. In each panel the lines represent  local thermal equilibrium (LTE) (blue), non-LTE  with temperature profile estimated by \citet{2006A&A...452..897C} (red) and non-LTE with temperature profile Infrared SED fitting by \citet{1995ApJ...439..288O}($T=25$\,K at $40$\,AU, green line). Given that the three models are identical in all other respects, the LTE assumption indeed results in a lower integrated flux densities.}
\label{LTEdisc}
\end{figure*}

\section{Results of synthetic observations}
\label{synthobs}

\subsection{Line profiles}

\begin{figure*}
\begin{center}
\includegraphics[width=\columnwidth]{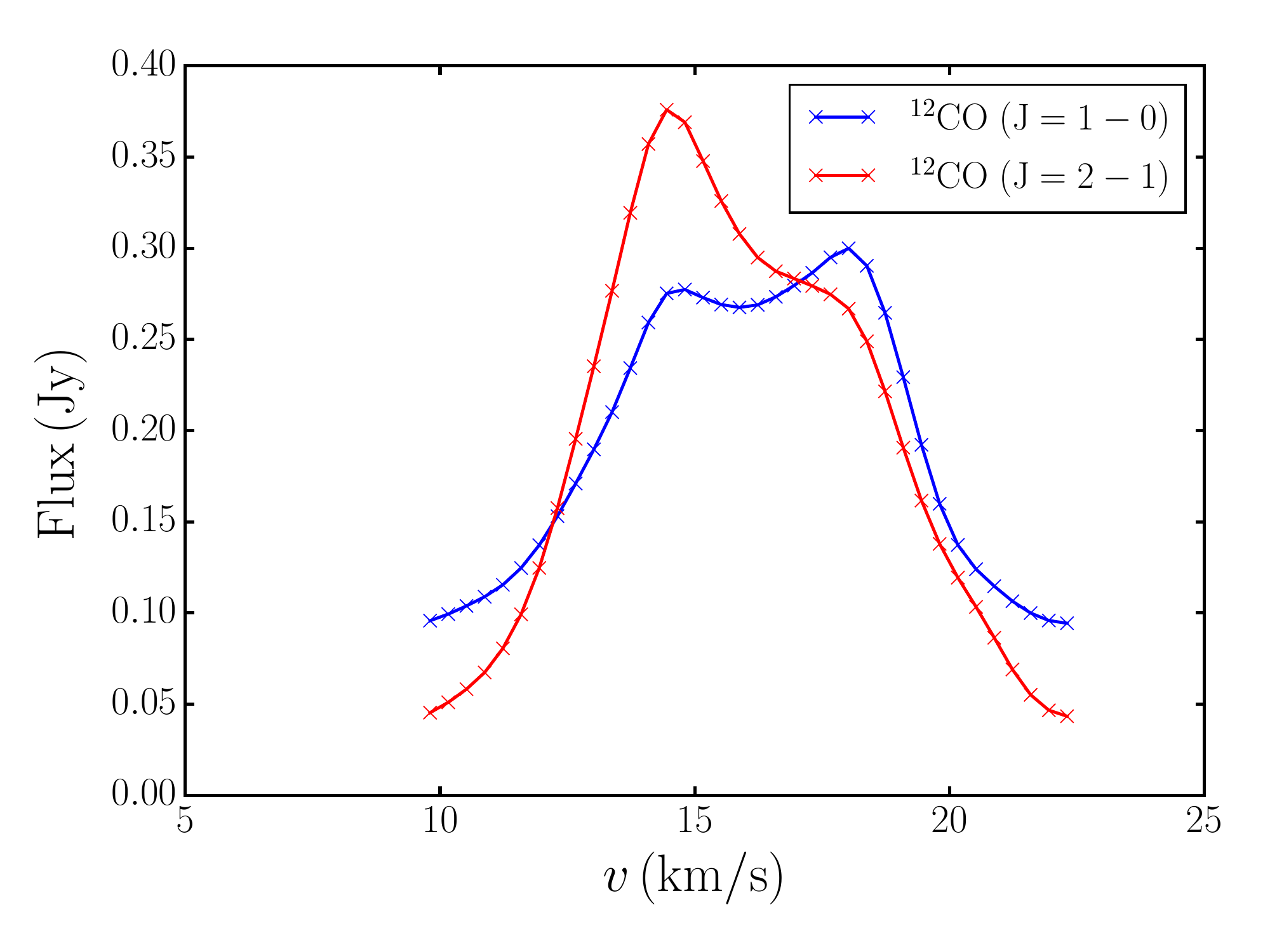}
\includegraphics[width=\columnwidth]{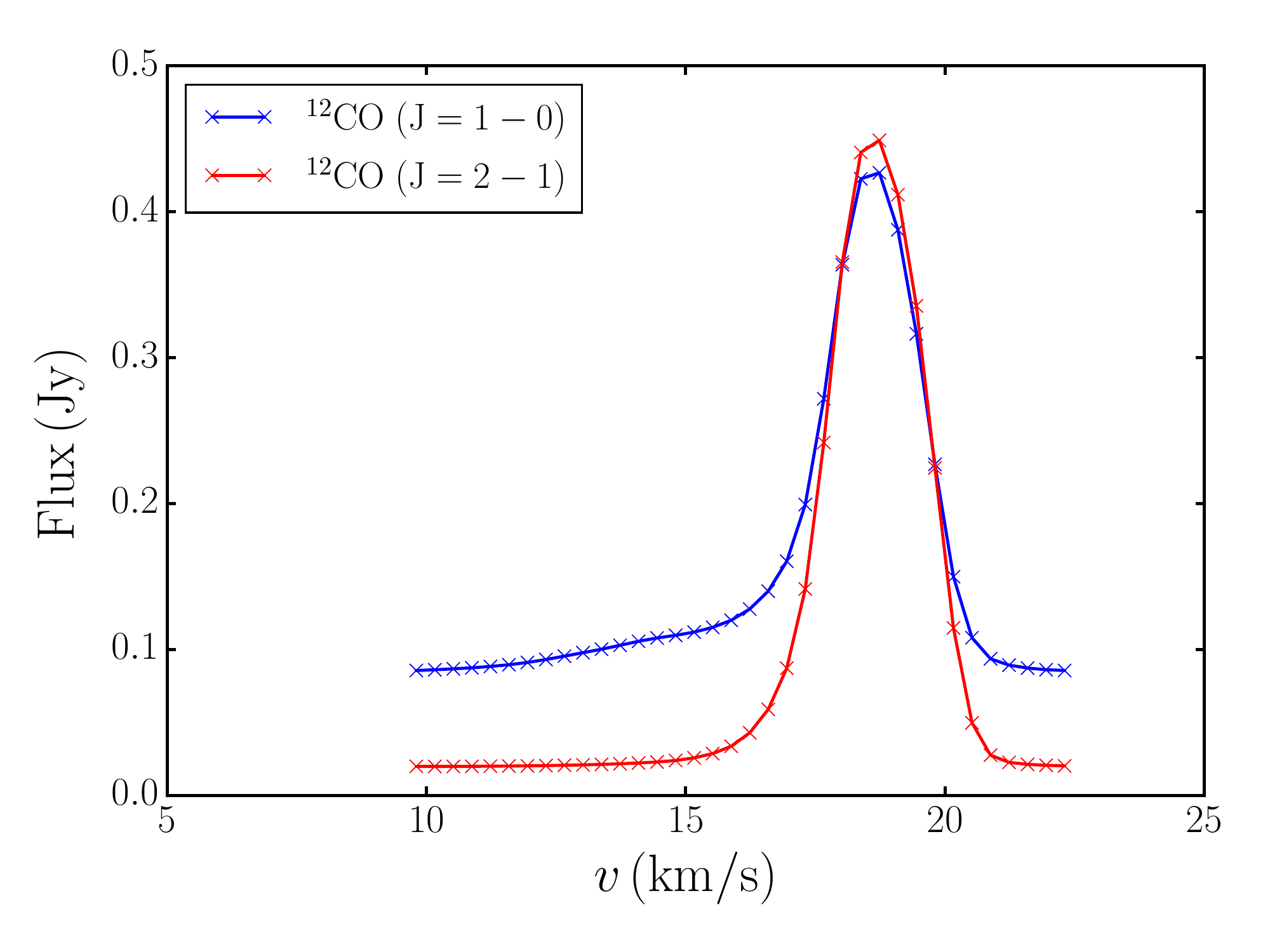}\\
\includegraphics[width=\columnwidth]{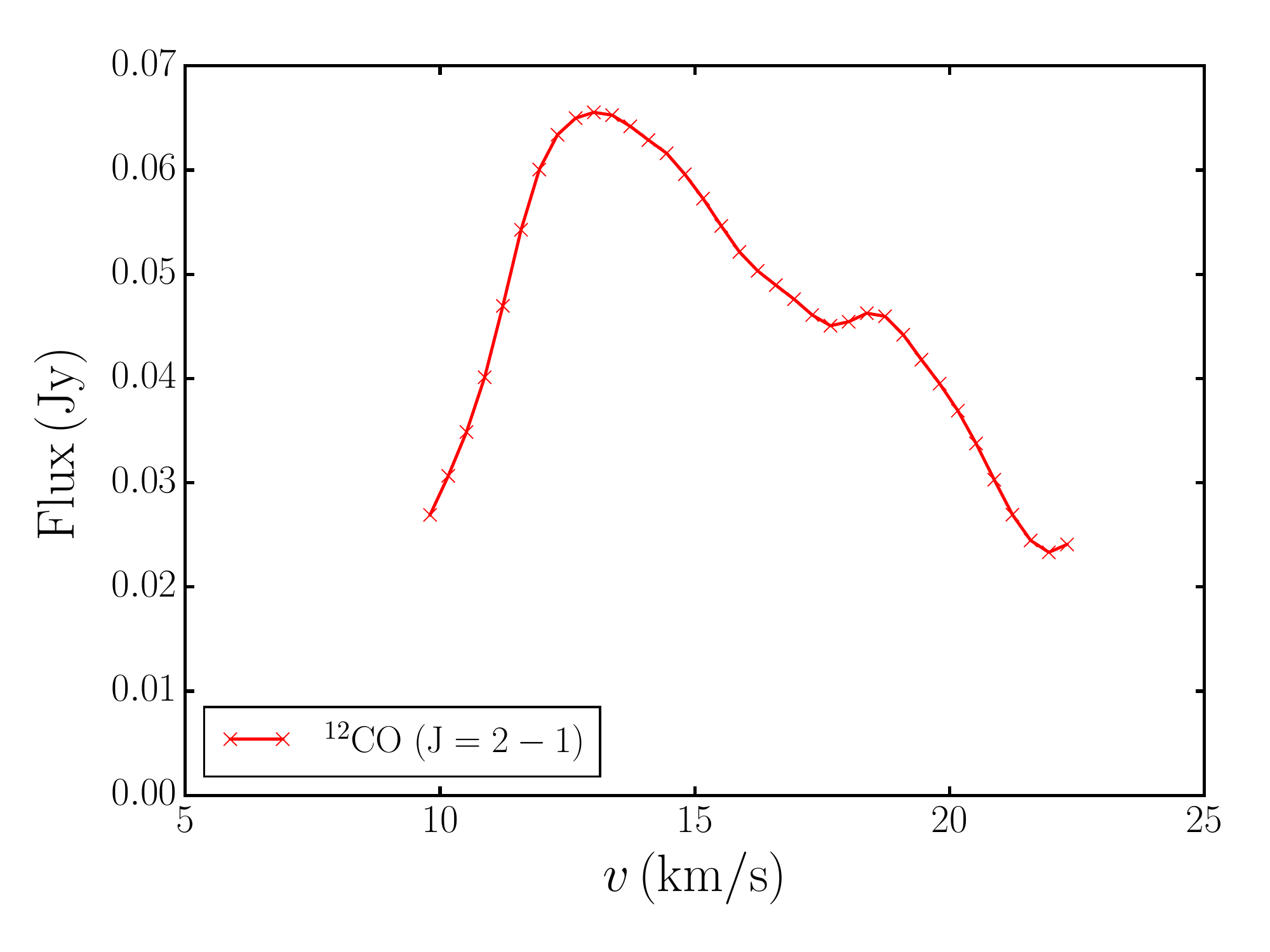}
\end{center}
\caption{Line profiles of the synthetic observations of the best fit model for the disc around star A (top left panel), the tidal arm (top right panel) and molecular gas around star B (bottom panel). The red curve shows the $^{12}$CO$(J=2-1)$ profile, and the blue curve the $^{12}$CO$(J=1-0)$ one. The latter has been multiplied by a factor $4$. The temperature profile is given by $T(R)=230\,{K}\,(R/{\rm AU})^{-0.5}$, and the mass of the disc is $M_{\rm d}=1.67\times10^{-3}M_\odot$. The synthetic profiles recover most of the features of the profiles derived by \citet{2006A&A...452..897C}, in particular flux intensities and kinematic signatures. The synthetic profile of the gas around star B is lower than the observed one by a factor of $2$. This is due to the simple prescription we have used for the temperature radial profile, which fails to reproduce the emission properties of the gas close to star B, where the heating due to radiation from star B is certainly important.}
\label{fig:line_profiles}
\end{figure*}

We took snapshots from our hydrodynamic models at $\sim570$ yr ($t_{\rm obs}\sim1.1P_{\rm peri}$, where $P_{\rm peri}$ is the Keplerian period at the pericentre) after pericentre and post processed them with \textsc{torus} to produce synthetic $^{12}$CO data cubes (c.f. section \ref{torussec}). Fig. \ref{fig:line_profiles} shows line profiles from subsets of  these data cubes after applying convolution accounting for the IRAM beam size, for the best-fitting model.

The synthetic observations allow us to constrain the temperature normalization and the initial mass of the disc $M_{\rm d}$, since these two quantities regulate the flux density normalization and the optical depth of the emission lines. In the case that the line is optically thick, the ratio $^{12}$CO$(J=2-1)$/ $^{12}$CO$(J=1-0)$ should be $\sim4$ in Rayleigh-Jeans limit. We have therefore multiplied the $^{12}$CO$(J=1-0)$ by $4$ for the regions of the tidal arm and the primary disc, to demonstrate whether these components of the system are optically thick.  We do not consider the $^{12}$CO$(J=1-0)$ line for the gas around star B, since this line has been marginally detected by \citet{2006A&A...452..897C}, as discussed above. The radial velocities were corrected to heliocentric frame by adding the mean heliocentric velocity of star A \citep[$15.87\pm0.55$\,km/s,][]{2006A&A...452..897C}. 

First, we consider the impact of temperature normalization. By using the temperature profile as estimated by \citet{2006A&A...452..897C} $(T(R)=450\,{\rm K}\times(R/{\rm AU})^{-0.5})$, the most outstanding discrepancy is that the simulated flux
densities are $3-5$ times higher than the observed flux densities.
We argue that the most likely cause is that the limitation of BS93 model led to \citet{2006A&A...452..897C} overestimating the temperature of
the disc. Although BS93 assumed local thermal equilibrium (LTE), 
this  assumption may not be appropriate for  RW Aur.
\textsc{torus} is capable of treating both LTE and non-LTE, allowing us to test this. The comparison in  
Fig. \ref{LTEdisc} demonstrates  that the resultant flux densities are 
indeed about $2-3$ times lower in the LTE case.

This indicates that in order to explain the observed flux, \citet{2006A&A...452..897C} would have overestimated 
the temperature by assuming an LTE model. Furthermore, if we use  a lower
temperature, more consistent with theoretical expectation from stellar irradiation and accretion, \citep[$T=30$\,K at $40$\,AU,][]{1997ApJ...490..368C}, the resultant theoretical emission is more compatible with the observed
 integrated flux densities. Note that  lower temperatures  are  also indicated by Infrared SED fitting by \citet{1995ApJ...439..288O}($T=25$\,K at $40$\,AU).

 We try to constrain the disc mass by requiring the tidal arm and the disc orbiting star A to be optically thick. This  requires  an
initial disc mass of  $1.67\times10^{-3}M_\odot$ and a current
 disc mass of $1.05\times10^{-3}M_\odot$. In order to estimate the latter, we have considered an outer disc radius of $57$\,AU, as estimated by \citet{2006A&A...452..897C}. This estimate is higher than that estimated by both \citet{2006A&A...452..897C} and \citet{1995ApJ...439..288O} ($10^{-4}M_\odot$ and $3 \times 10^{-4}M_\odot$, respectively), but lower than the ones estimated by \citet{2005ApJ...631.1134A} ($4\times10^{-3}M_\odot$). Note however that mass estimates from dust and CO isotopologues are always affected by very large uncertainties, and that our prescription for the temperature dependence is so simple that we do not expect to recover the observed mass with high precision.
 
After adopting the temperature normalization and initial disc mass as described above, the resultant line profiles show better agreement with the observations. In particular, both the A disc and the tidal arm are optically thick as observed, and  their peak flux is in broad agreement with the observations by \citet{2006A&A...452..897C}, together with the moderate asymmetry of the double-horned line profile of the disc. The disc has in fact gained a small eccentricity, as apparent from the right bottom panel of Fig. \ref{fig:sph}. The $^{12}$CO$(J=2-1)$ line profile of the gas around star B recovers the main features of the observed profile, besides the peak flux, which is lower than the observed one by a factor of $2$. This may result from incorrect assumptions
about the temperature profile around star B (temperature has a simple dependence on the distance from star A), but may also simply reflect that
star B also possessed a disc prior to the encounter (in contrast to
what we assume here, where the final disc is composed of material
entirely captured from the disc of star A).

\begin{figure}
\includegraphics[width = \columnwidth]{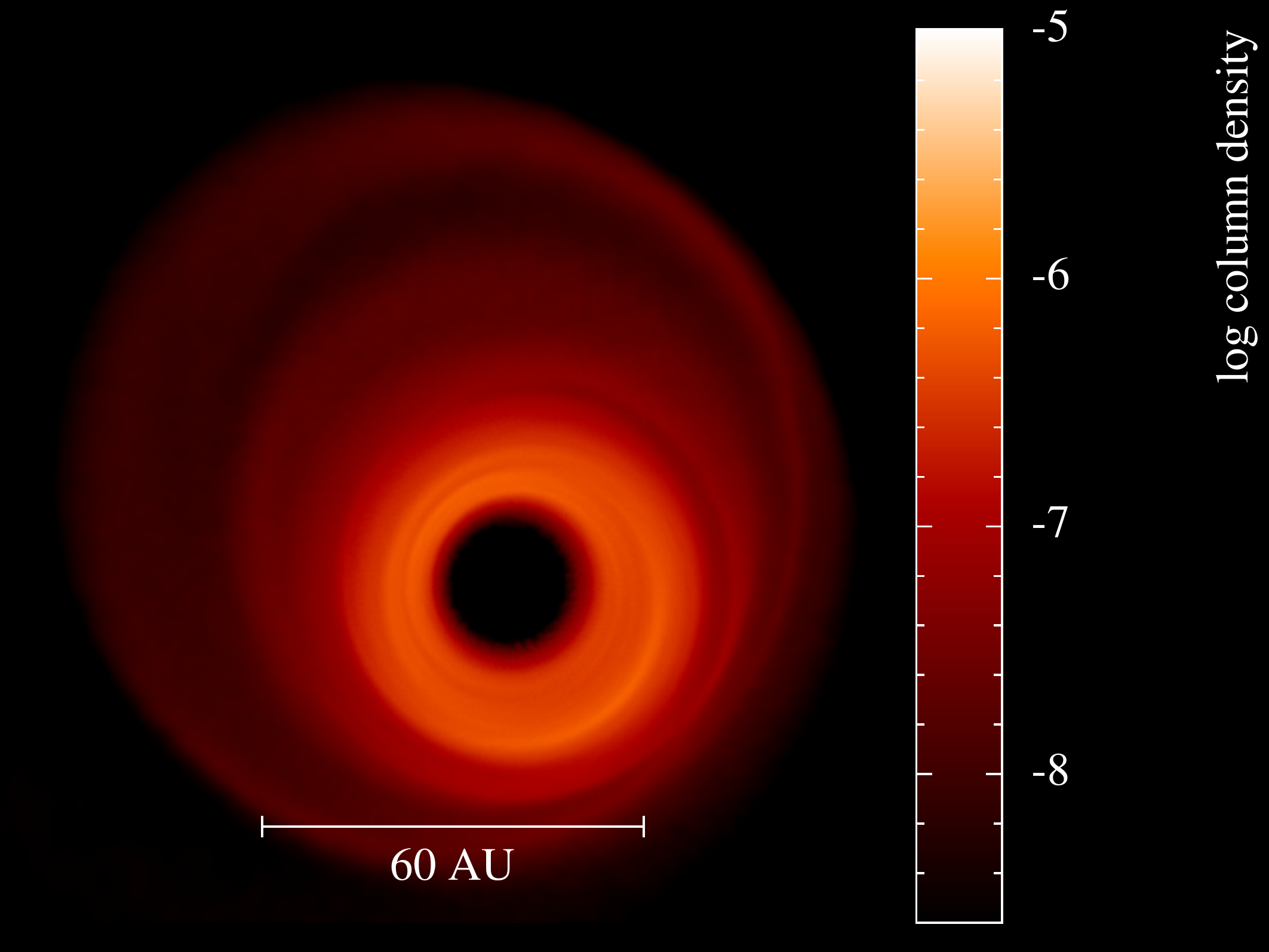}
\caption{Circumstellar disc around star A as seen at $t=0$ (today) from the SPH simulation on the $xy$-plane. Column density is in code units. The outer radius of the disc is in agreement with values estimated by \citet{2006A&A...452..897C} ($40 - 57$\,AU). The disc has clearly gained an eccentricity after the encounter with star B.}
\label{fig:disc}
\end{figure}

\begin{figure*}
\begin{center}
\includegraphics[width = \columnwidth]{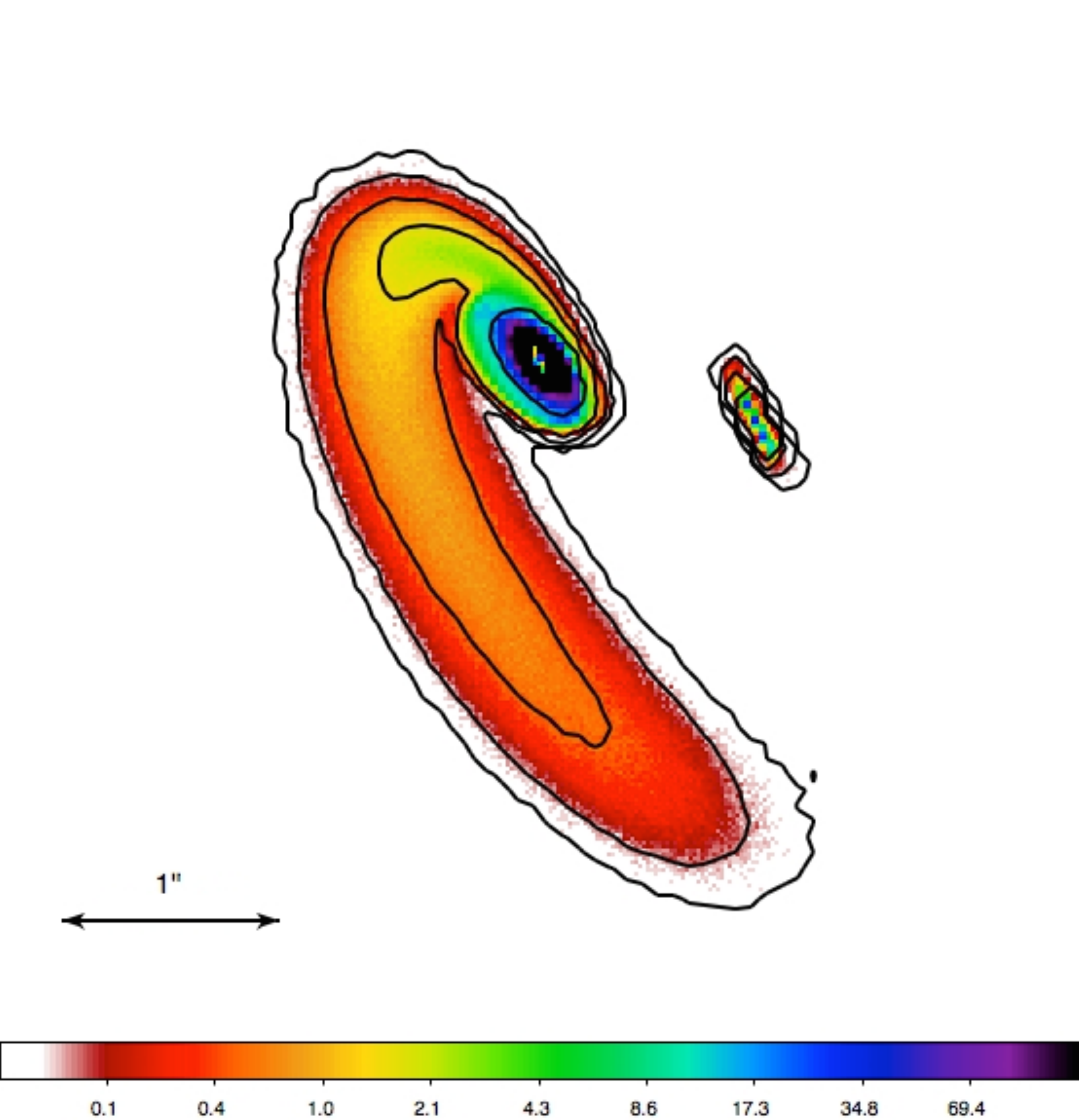}
\includegraphics[width = \columnwidth]{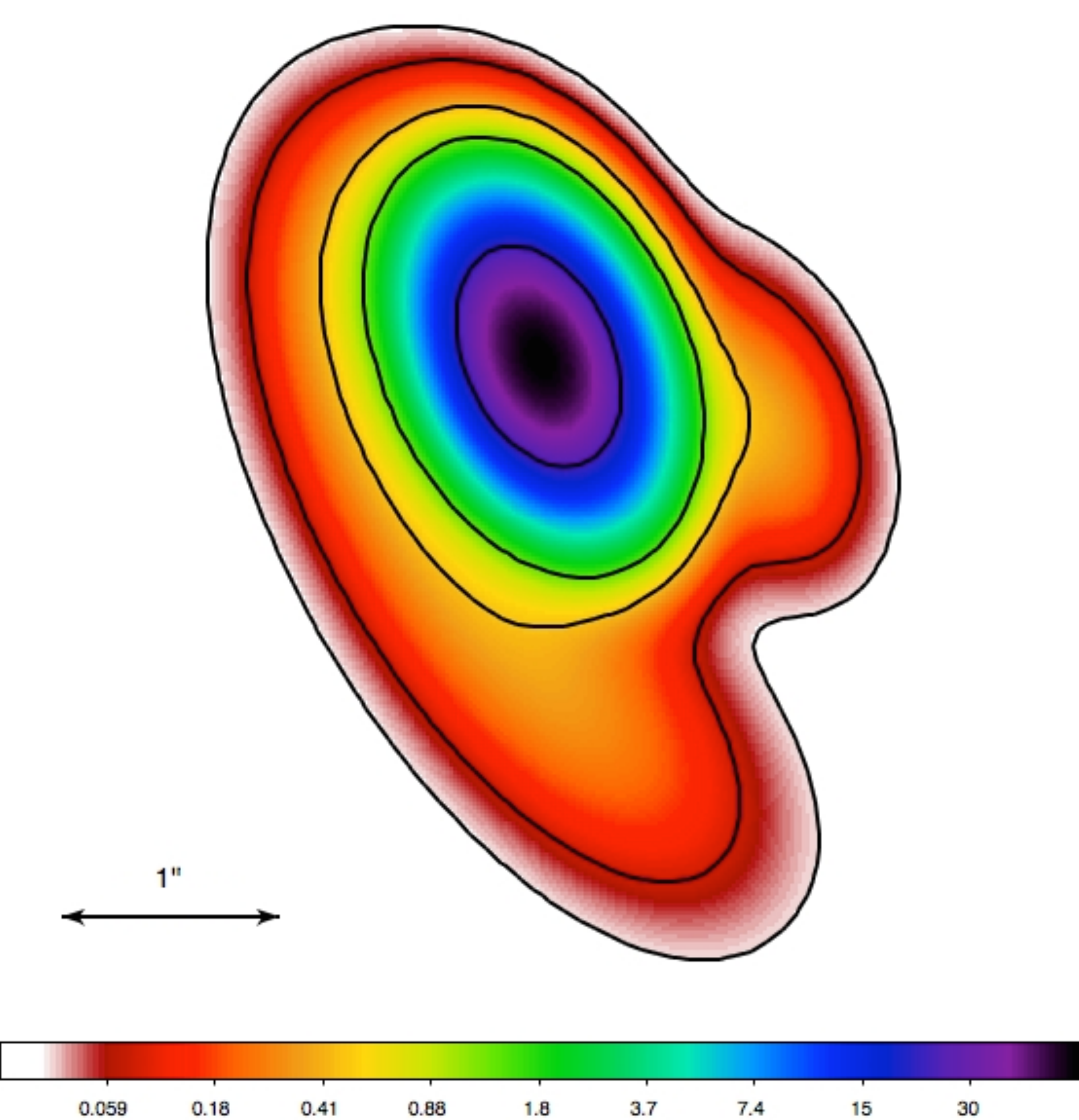}
\end{center}
\caption{Left panel: synthetic observation of our best-fit model in the continuum at $1.3$\,mm. Right panel: same emission map, convolved with the beam size used by \citet{2006A&A...452..897C} (FWHM: $0.87\arcsec\times0.54\arcsec$; PA $=20^\circ$). The colour code shows emission in mJy/beam. Contour levels are at (1/24, 1/6, 1, 3 and 48)$\sigma$, where $\sigma=0.47$\,mJy/beam \citep[see Table 1 by][]{2006A&A...452..897C}.}
\label{fig:dust}
\end{figure*}

\citet{2006A&A...452..897C} obtained an indirect measure of the size of the disc around star A by using the position of the peaks of the typical double-horned profile of the emission lines (see Section \ref{sec:star_a}). They observed a line profile with peaks at $\sim3.5$\,km/s from the centre, whereas in our synthetic observation we obtain the two peaks at $\sim2.5$\,km/s from the centre of the line. This seems to suggest that the disc in our simulations is larger than the one observed by a factor $\sim2$. However, if we look at the hydrodynamical simulation of our best model (see Fig. \ref{fig:disc}), or if we look at the synthetic continuum emission shown in Fig.  \ref{fig:dust}, the disc size is in good agreement with the observations ($\sim50$\,AU). After convolution we are obtaining contamination from low velocity gas in the inner part of the tidal arm and the elliptical outer region of the disc. The difference in our synthetic line profiles implies
that we are over-estimating emission from these outer regions. Detailed
thermal modelling is however beyond the scope of this project;
the agreement with the size derived from the thermal continuum reassures
us that the dynamical effects of the encounter are well reproduced.

\subsection{Dust emission}
\label{dust}
We used \textsc{torus} (c.f. section \ref{torussec}) to produce a synthetic observation of the continuum emission at 1.3\,mm of the same snapshot showed in Fig. \ref{1to0}. We consider standard assumptions on the dust properties. First of all we assume that spatial distribution of the dust follows exactly the gas distribution. This is likely to be the case in our framework, since the dynamics of the gas itself is mainly ballistic during the encounter. The dust component is dynamically driven by the same gravitational potential. At the moment we neglect secondary effects, that might be due to hydro terms in the dynamics of the gas. We assume a gas to dust ratio equal to $100$, and a grain size distribution ranging between $50$\,\angstrom\ and $1$\,cm. We use a typical grain size distribution, where $dn/ds \propto s^{-w}$, where $n$ is the numerical density of dust particles, and $s$ is the grain size. It is well known that in the interstellar medium $w$ assumes a typical value $w \sim 3.5$ \citep{mathis_77}. More recently, interferometric observations have suggested a shallower distribution in the outer regions of protoplanetary discs, where $w \sim 3$ \citep[e.g.][]{2010A&A...521A..66R,2010A&A...512A..15R,2012A&A...540A...6R,2014arXiv1402.1354T}. Following these recent results, for our synthetic observation we use $w=3.3$. We assume that the dust is comprised of amorphous silicate grains \citep{2003ApJ...598.1017D,2003ApJ...598.1026D}, obtaining an opacity $\kappa_{\nu} = 0.131$\,cm$^2$/g at $1.3$\,mm (assuming a gas to dust ratio equal to $100$). Fig. \ref{fig:dust} shows the emission we obtain in mJy/beam, where the beam size is the one used by \citet{2006A&A...452..897C} ($0.87\arcsec\times0.56\arcsec$, see their Table 1). Contour levels are submultiples and multiples of the noise level $\sigma=0.47$\,mJy/beam in the same observation by \citet{2006A&A...452..897C}: (1/24, 1/6, 1, 3, 48)$\sigma$. From disc A we obtain emission levels that are compatible with the one observed (given the large uncertainties on the opacity), and a disc size that is in agreement with the one estimated by \citet{2006A&A...452..897C} from the line profiles ($\sim 50$\,AU from the unconvolved emission map). We underestimate the dust emission around star B, and this is again suggesting that our simple temperature treatment is underestimating the temperature of the material around the secondary star. By comparing the convolved emission map with the one shown by \citet{2006A&A...452..897C} where the primary disc is located, it seems that we are overestimating the total emission by a very small factor $\sim1.5$ (compare the size of the $3\sigma$ contours), which is very small when compared to the uncertainties of the problem. Note that the normalization of the emission of our synthetic observation does depend on our assumptions, mostly the dust opacity \citep[e.g. in discs is often assumed a lower opacity at 1.3\,mm, as motivated by][]{1990AJ.....99..924B}, but also the uniform dust-to-gas ratio, the maximum grain size, and the grain size distribution index. With the assumptions we have made, the emission from the tidal tail is below the noise level ($\sim0.3$\,mJy/beam).

\section{The dimming event}
\label{sec:dimming}

\begin{figure*}
\begin{center}
\includegraphics[width=\columnwidth]{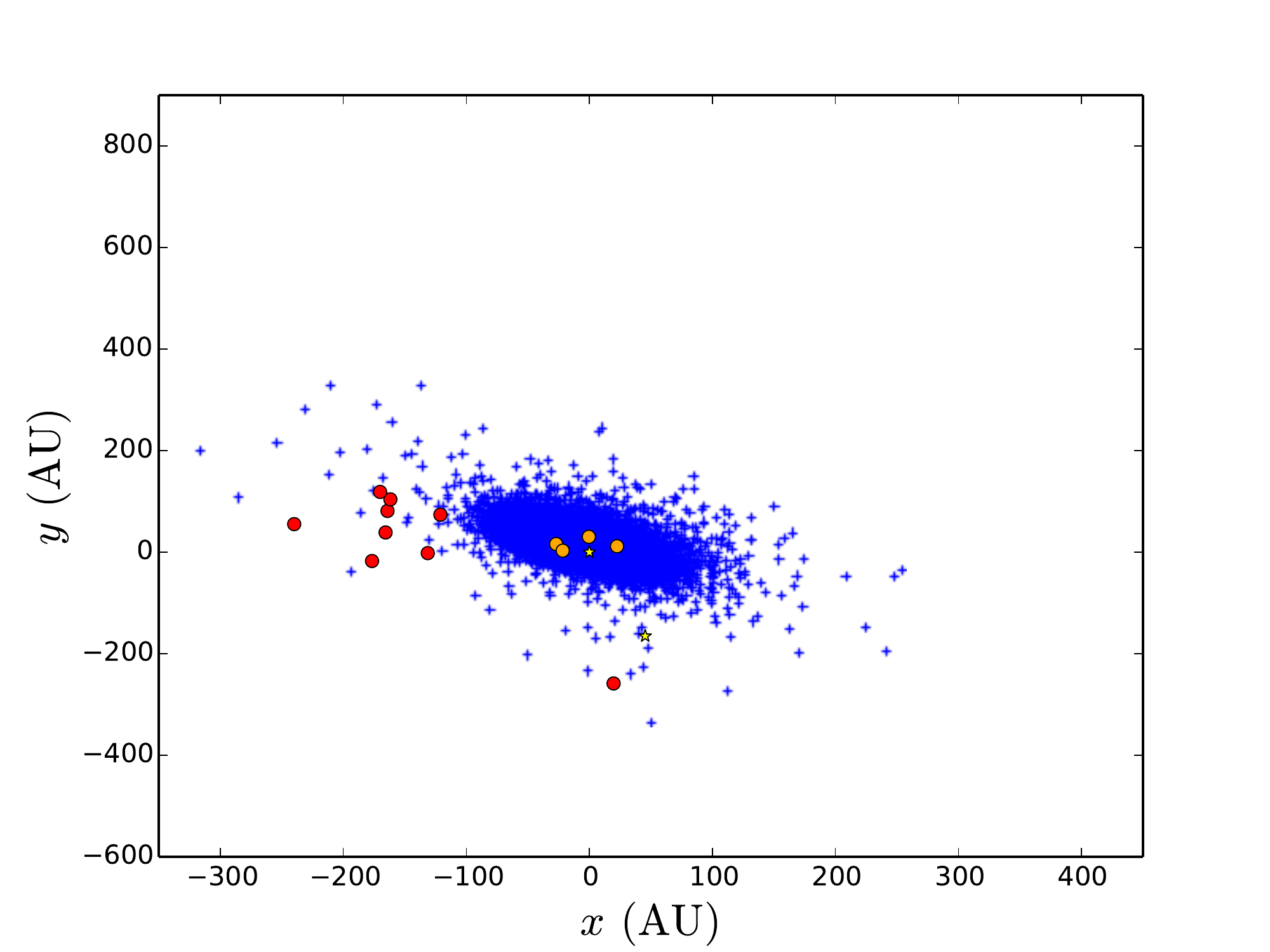}
\includegraphics[width=\columnwidth]{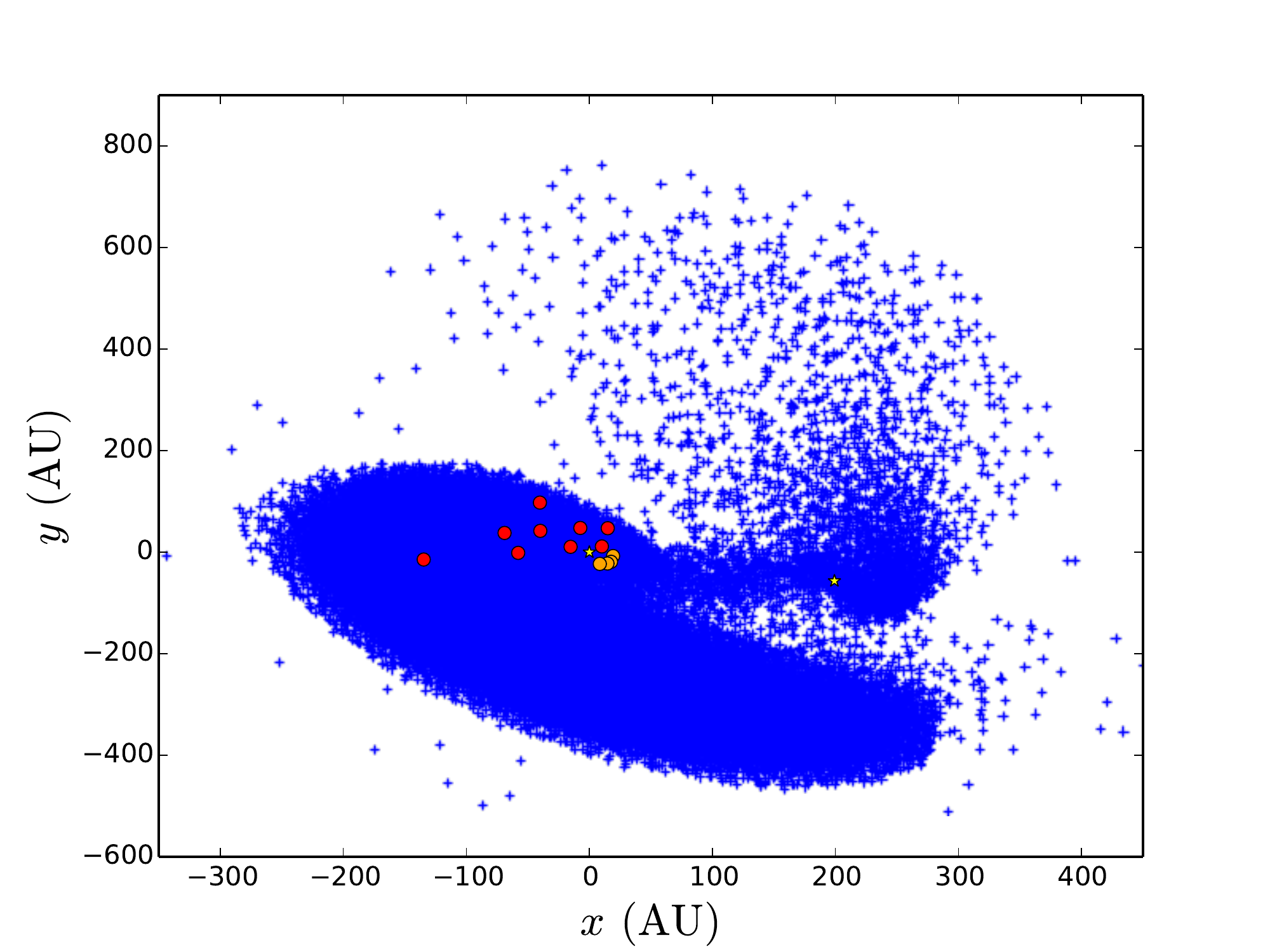}\\
\includegraphics[width=\columnwidth]{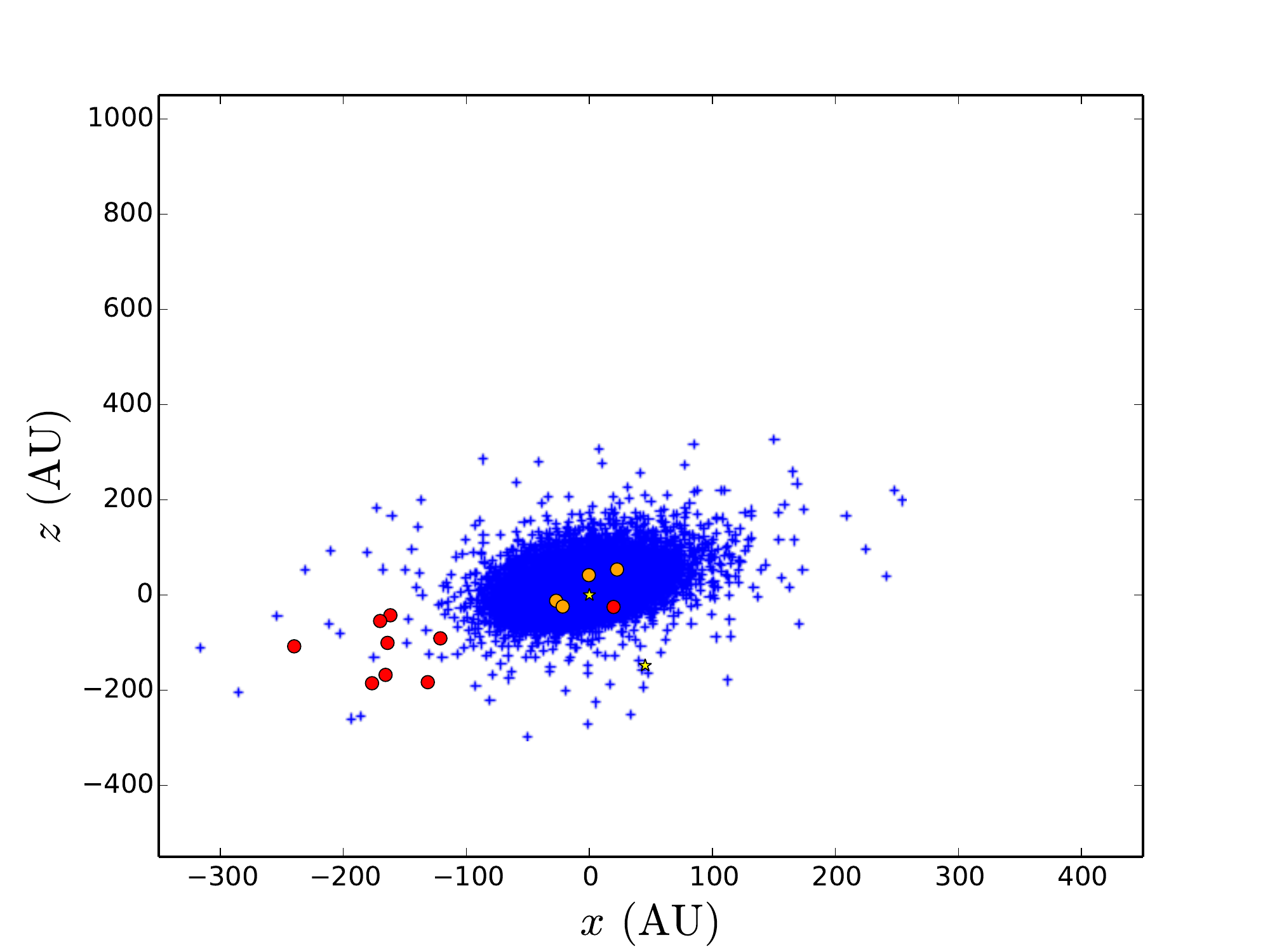}
\includegraphics[width=\columnwidth]{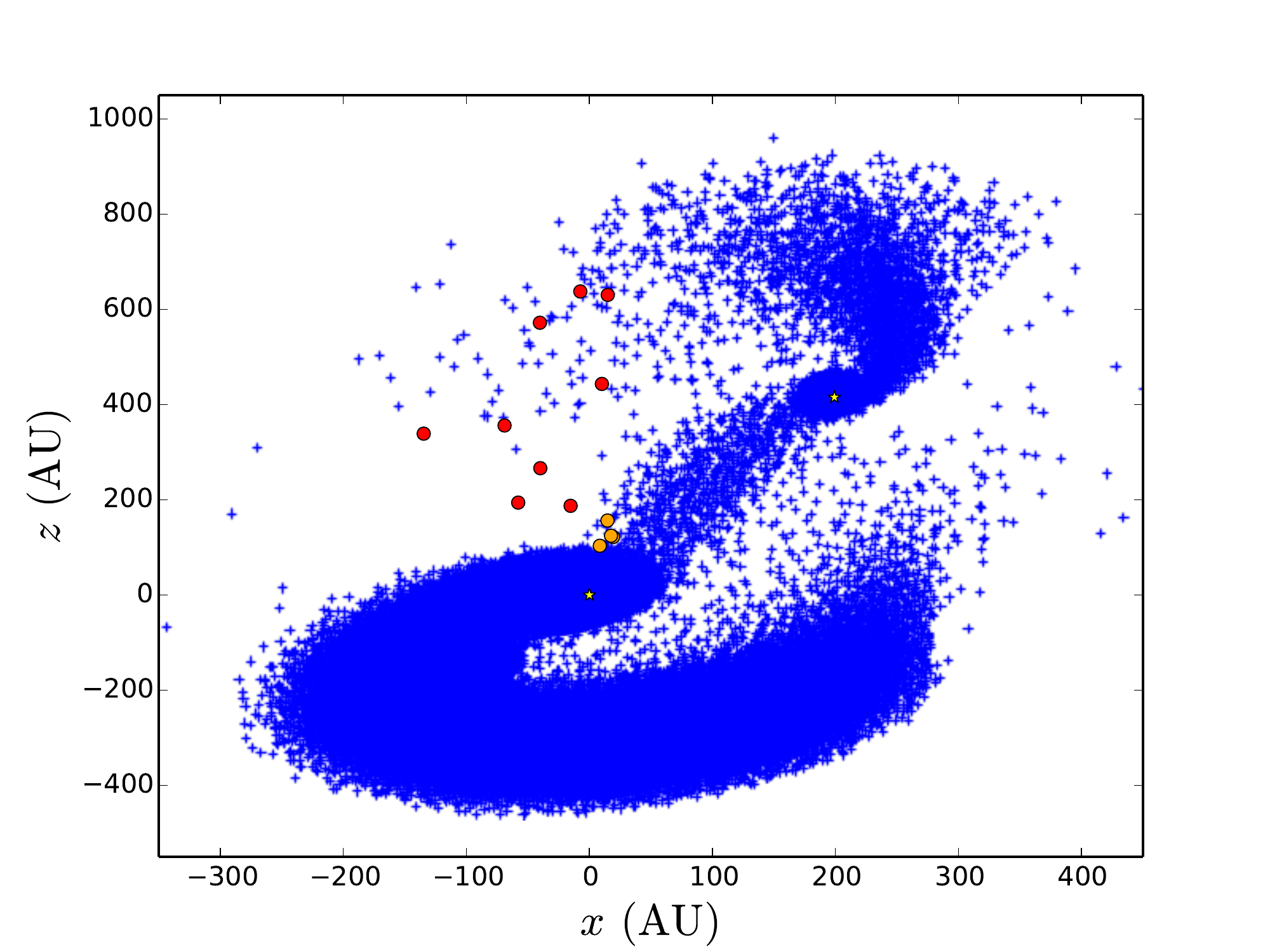}\\
\includegraphics[width=\columnwidth]{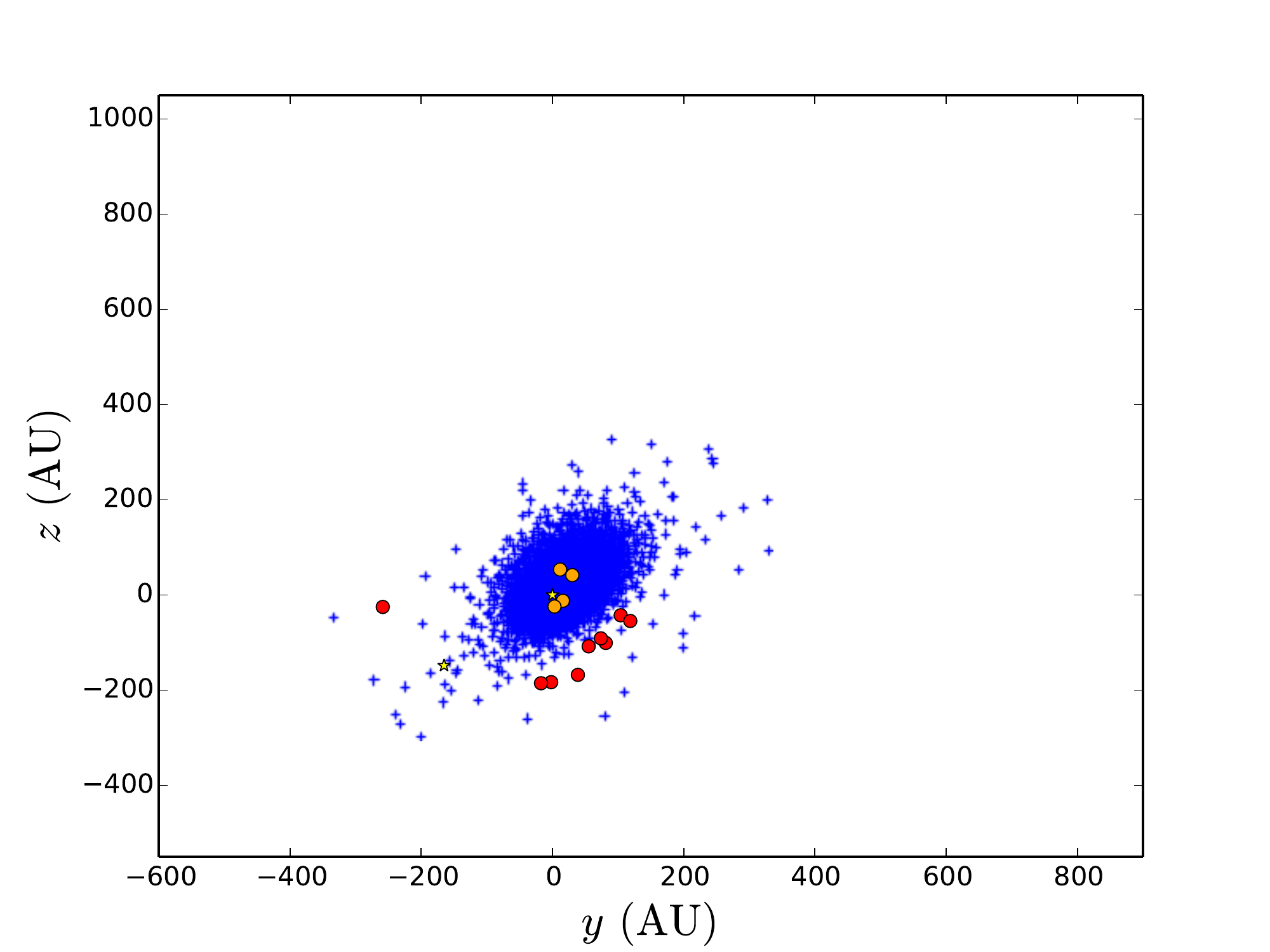}
\includegraphics[width=\columnwidth]{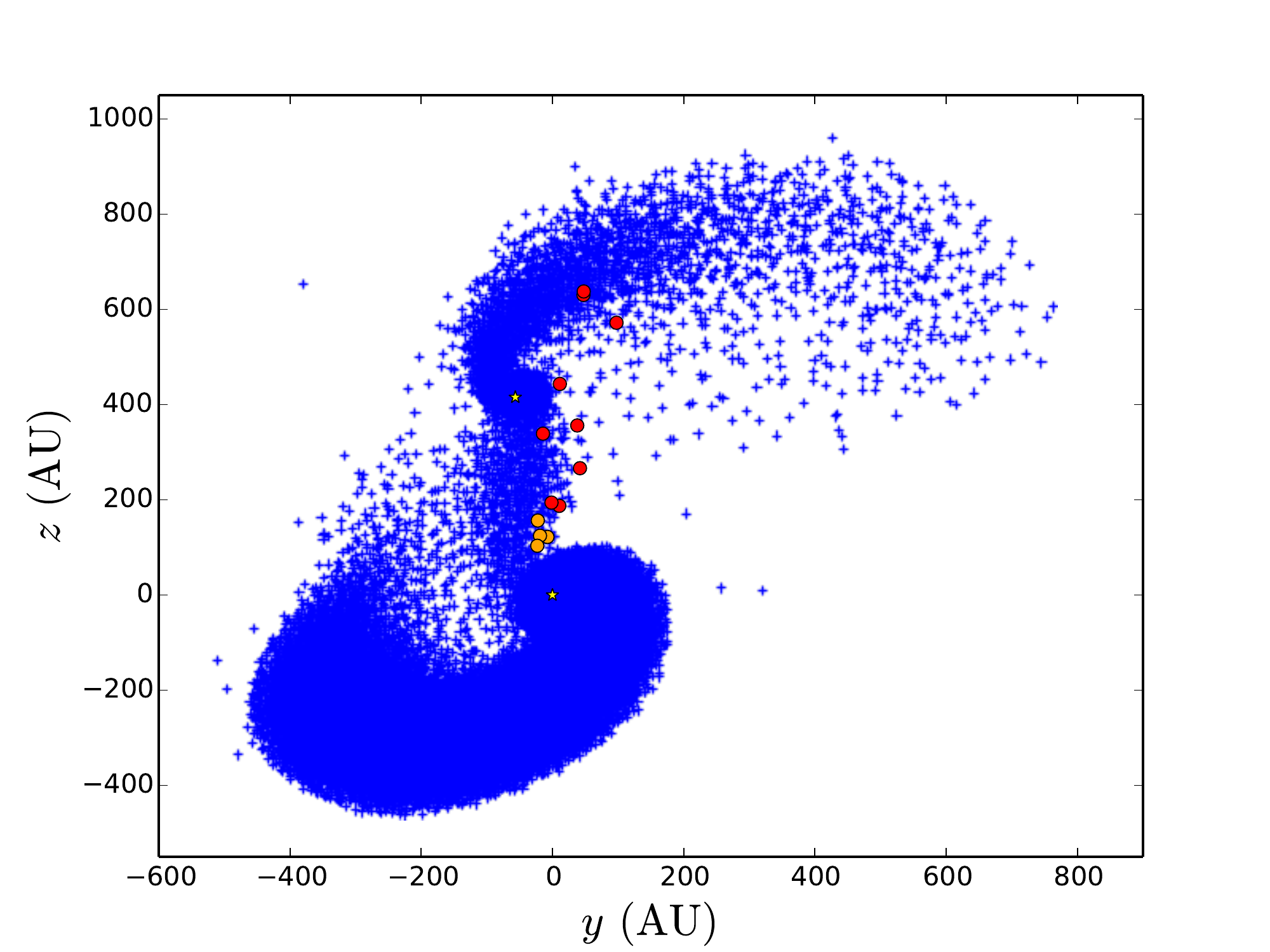}
\end{center}
\caption{3D structure of RW Aur before (left panels, at $-636$ yr from today) and after (right panels, at $0$ yr from today) the encounter for the best-fitting model. As in the other figures, the $x$-$y$ plane represents the plane of the sky. The observer is located at $z=+\infty$. The origin of the axes is defined as the position of star A. Blue crosses represent all the SPH particles that have not been accreted on either of the two stars, which are shown in yellow. Filled dots indicate the particles that are partially occulting star A at present time. The red ones are scattered particles that do not show a credible dynamical history. The orange ones evolve in a coherent fluid structure. They are initially orbiting within the disc, are then stripped by the tidal interaction, and end up in the low density periphery of the stream connecting stars A and B. The dimming event observed by \citet{2013AJ....146..112R} can be explained by a clumpy inhomogeneity at the edge of such stream. Transverse velocities of the particles and column density are in broad agreement with the estimates by \citet{2013AJ....146..112R}.}
\label{fig:occultation}
\end{figure*}

\citet{2013AJ....146..112R} argued that the leading edge of the tidal arm may wrap around the A disc and obscure the line of sight between Earth and RW Aur A, thereby producing  the observed dimming of RW Aur. In our simulations, there is indeed a stream of gas linking star A and B  after the tidal encounter \citep[however, it is not part of the primary tidal arm as][suspected, see the bottom left panel of Fig. \ref{fig:sph}]{2013AJ....146..112R}. A portion of this stream of gas might provide  the occulting body that caused the observed dimming event, as the integrated line-of-sight column density ($\sim 10^{-2}$\,g\,cm$^{-2}$) exceeds the threshold column density ($\sim 7\times10^{-4}$\,g\,cm$^{-2}$) required to produce the dimming - assuming a flux decrease of the A star of $91\%$ \citep{2013AJ....146..112R} and a dust opacity at optical wavelengths $\kappa=3000$\,cm$^2/$g \citep{1997A&A...318..879M}.

In order to verify this hypothesis, we have analysed the SPH simulation of the best-fitting model. We select all the foreground particles that are partially occulting star A at the present time by requiring  $h/D>1$, where $h$ is the smoothing length associated with a single SPH single particle and  $D$ is the distance between the particle and star A in the plane of the sky.  We then trace back their trajectories, in order to see whether the selected particles have a credible dynamical history. In Fig. \ref{fig:occultation} we plot all the SPH particles of the simulation with blue crosses, the two stars with yellow star symbols, and the occulting particles with filled dots. The orange dots represent the particles that show a credible dynamical history. They start orbiting together within the circumstellar disc (see left-hand panels), and they then get incorporated in the well resolved bridge linking star A and B. At the present time they are located at the low density periphery of the bridge (see right-hand panels). The occulting particles represented by the red dots are individually scattered particles that originate from the very poorly resolved outer regions of the initial disc and are not associated with any coherent fluid structure. Since the orange particles are very few (only 4 for the particular choice of viewing angle), we cannot determine the structure of the low density margins of the bridge, and even less the size of the occulting body \citep[$0.09-0.27$\,AU, as estimated by][]{2013AJ....146..112R}. An accurate determination of the density structure of the low density periphery
of this bridge would require an unfeasibly high resolution. However, the
presence of particles in this region  indeed indicates  that possible 
$\sim$ AU scale inhomogeneities at the edge of the stream linking A and B 
could cause the  dimming event observed by \citet{2013AJ....146..112R}. The velocity in the plane of the sky of these particles ranges between $0.97-1.09$\,km$/$s and is thus consistent with
the estimates from \citet{2013AJ....146..112R} ($0.86-2.58$\,km$/$s).

\begin{figure}
\includegraphics[width = \columnwidth]{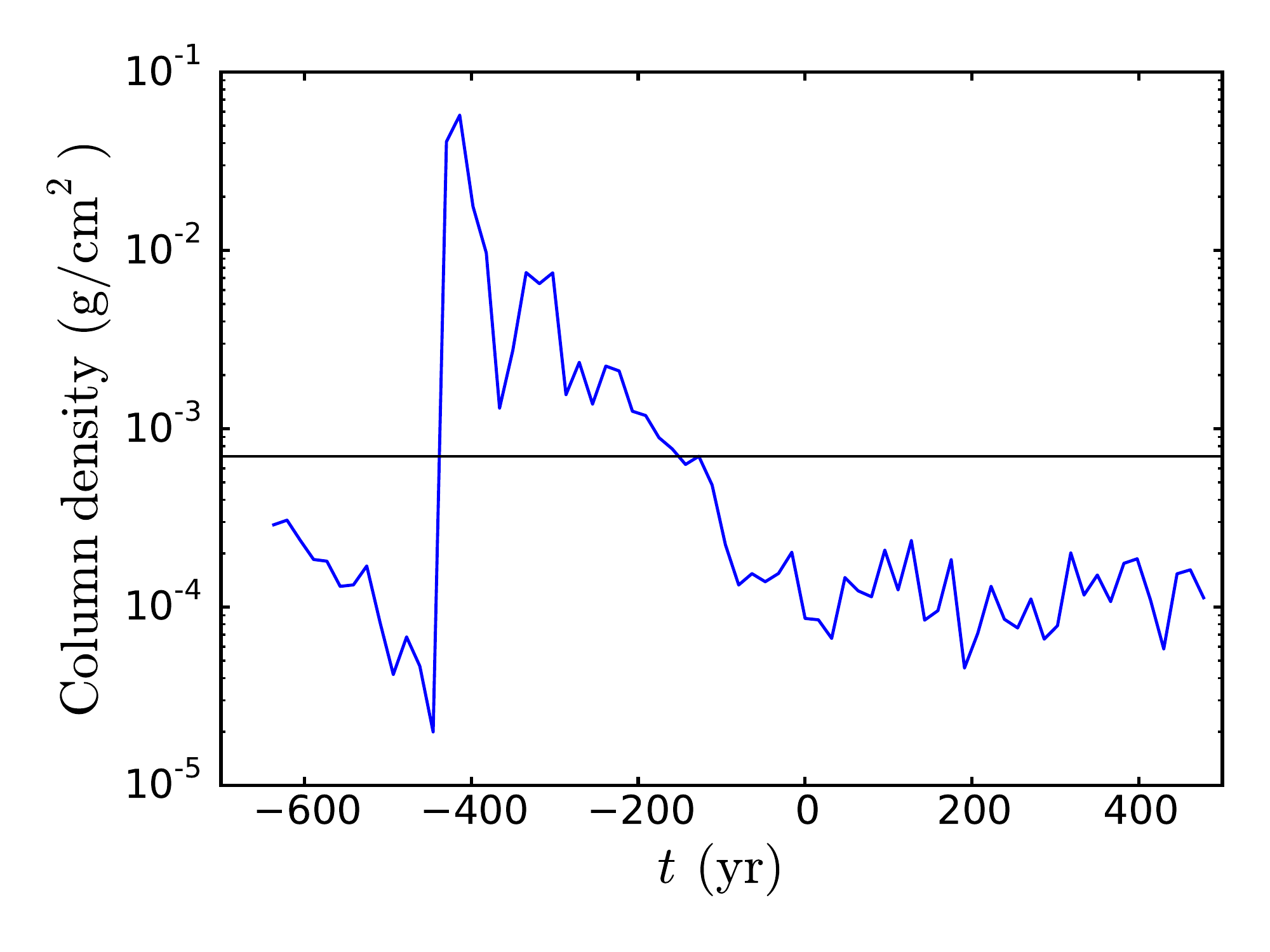}
\caption{Average column density between the observer and star A as a function of time. The horizontal black line represents the column density that would cause a dimming of $\sim90\%$, as observed by \citet{2013AJ....146..112R}. In the past the bridge connecting stars A and B has been perfectly on the line of sight of star A, therefore completely occulting the star at optical wavelengths. Nowadays such structure grazes the line of sight. Future dimming events cause by inhomogeneities in the density structure can not be ruled out by this model.}
\label{fig:col_dens}
\end{figure}

Finally, we have computed the evolution with time of the average column density lying between star A and the observer. In Fig. \ref{fig:col_dens} we show that in the past the star has been completely occulted (at optical wavelengths) by the density structure linking stars A and B. This bridge has then evolved, until the line of sight only grazes it nowadays ($t=0$). In the future, the average column density stays well below the value required to observe a dimming of star A as observed by \citet{2013AJ....146..112R}. Moreover, for $t\gtrsim0$, the column density in the simulation starts to be dominated by particle noise. However, inhomogeneities in the bridge structure might entail new events in the next years.

\section{Summary and conclusions}

\label{sec:concl}

We have used hydrodynamic models and synthetic observations to constrain the observed morphology, kinematics and line fluxes  of the binary system RW Aur which was first proposed as a candidate star-disc encounter by \citet{2006A&A...452..897C}. We draw the following conclusions from this work. \\

 With a suitable set of orbital parameters and viewing angle
(See Table 1), we can reproduce almost all morphological and
dynamical features of the system:
\begin{itemize}
\item{RW Aur A has a circumstellar disc with an outer radius of $\sim50$AU which is within the observed range of $40-57$AU. The disc is optically thick in $^{12}$CO
emission, as the $^{12}$CO$(J=2-1)$/ $^{12}$CO$(J=1-0)$ ratio is $\sim4$.}

\item{RW Aur B is separated from star A by  $\sim200$\,AU ( $\sim1.5\arcsec$ for
$140$ pc). The projected AB separation is increasing at rate of
$\sim0.002\arcsec$/yr. The molecular complex on B is blueshifted with
respect to star A with a broad distribution of radial velocity of
$3-6$\,km/s. The line profile from the star B is asymmetric and
thus most likely due to material captured during tidal
encounter instead of a circumstellar disc. The molecular
emission is marginally optically thick with a  $^{12}$CO$(J=2-
1)$/ $^{12}$CO$(J=1-0)>4$.}

\item{The expanding, spiral-shaped structure connected to RW
Aur A disc is most likely a tidal arm produced in a fly-by of
star B. The tidal arm extends about $600$\,AU ( $\sim4\arcsec$ for $140$\,pc).
The arm is entirely redshifted with a radial velocity  $\sim3$\,km/s
with respect to star A. This is faster than the escape velocity at
corresponding distance, indicating the arm is indeed
gravitational unbound.}
\end{itemize}

The synthetic molecular line data cubes obtained from a purely
radial temperature estimated by \citet{2006A&A...452..897C} have flux
densities $3-5$ times higher than observations. Nonetheless, by
lowering the temperature to typical disc temperature for CTTS,
we can  reduce the flux densities to the observed values.

The agreement between observations  and simulations  thus lends strong support to 
the hypothesis that the morphology of the RW Aur system
is consistent with a tidal encounter scenario. Although the observational
evidence for tidal  encounters on Galactic scales is well
established, this is the
 only system that is a good candidate for such encounters on the scale
of protoplanetary discs \citep[note that a tidal encounter scenario has also been proposed as a possible model to explain the recent ALMA observations of AS 205 S, see][]{2014ApJ...792...68S}.

 Finally we note that our simulations provide  support for the
hypothesis of  \citet{2013AJ....146..112R} regarding the origin of observed
 dimming of star A in 2010/2011. We find that it is plausible that this
was caused by inhomogeneities in the outskirts of a `bridge' structure linking star A and B which may graze the line of sight at current epochs. Our
simulations do not have the resolution to trace the structure of the
low density periphery of the optically thick bridge, nor to model the
$\sim$ AU scale inhomogeneities that would be required to explain such
an event. Nevertheless our simulations predict a good agreement  between 
the velocity in the plane 
 of the sky of material in this region with that required by the observed ingress 
time of the dimming event. This provides good support for the occultation
hypothesis. It suggests that - depending on the filling factor of such
inhomogeneities - further dimming events should not be ruled out \citep[as confirmed by the observations by][]{2015IBVS.6126....1A} and that
these would be expected to share similar kinematic properties.

\section*{Acknowledgements}
We thank the anonymous referee, whose comments helped improving this manuscript. We thank Daniel Price for giving us the permission to use the \textsc{phantom} code, Joey Rodriguez and Keivan Stassun for a careful reading of the manuscript, and Sylvie Cabrit and Tim Harries for useful discussions. The synthetic observation calculations in this paper were performed using the Darwin Data Analytic system at the University of Cambridge, operated by the University of Cambridge High Performance Computing Service on behalf of the STFC DiRAC HPC Facility (www.dirac.ac.uk). This equipment was funded by a BIS National E--infrastructure capital grant (ST/K001590/1), STFC capital grants ST/H008861/1 and ST/H00887X/1, and DiRAC Operations grant ST/K00333X/1. DiRAC is part of the National E--Infrastructure. SF is funded by an STFC/Isaac Newton Trust studentship. Fig. \ref{fig:sph} was produced using \textsc{splash} \citep{price07}, a visualization tool for SPH data.
This work has been supported by the DISCSIM project, grant
agreement 341137 funded by the European Research Council under
ERC-2013-ADG.

\bibliographystyle{mn2e}
\bibliography{molecular}

\bsp

\label{lastpage}

\end{document}